\documentstyle[aps,prc]{revtex}
\def\shiftleft#1{#1\llap{#1\hskip 0.04em}}
\def\shiftdown#1{#1\llap{\lower.04ex\hbox{#1}}}
\def\thick#1{\shiftdown{\shiftleft{#1}}}
\def\b#1{\thick{\hbox{$#1$}}}
\begin{document}

\title{Moscow-type NN-potentials and three-nucleon bound states}
\author{ }

\maketitle

\begin{center}
V. I. Kukulin$^{1,2}$, V. N. Pomerantsev$^{1}$, 
Amand Faessler$^{2}$,  A. J. Buchmann$^{2}$, \\and E. M. Tursunov$^{3}$ 
\\[0.25cm]
$^1${\footnotesize{\it Institute of Nuclear Physics, Moscow State
University, 119899 Moscow, Russia}} \\
$^2${\footnotesize{\it Institute for Theoretical Physics,
University of T\"ubingen, Auf der Morgenstelle 14,\\
D-72076 T\"{u}bingen, Germany}} \\
$^3${\footnotesize{\it Institute of Nuclear Physics Academy of Sciences, 
Tashkent, Uzbekistan}}
\end{center}

\begin{abstract}
A detailed description of Moscow-type (M-type) potential models for the 
NN interaction is given. The microscopic foundation of these models, which
appear as a consequence of the composite quark structure of nucleons,
is discussed. M-type models are shown to arise naturally in a 
coupled channel approach when compound or bag-like six-quark 
states, strongly coupled to the NN channel, are eliminated
from the complete multiquark wave function. The role of the 
deep-lying bound states that appear in these models is elucidated. 
By introducing additional conditions of orthogonality to these compound 
six-quark states, a continuous series of almost on-shell equivalent 
nonlocal interaction models, characterized by a strong reduction or
full absence of a local repulsive core (M-type models), is generated.
The predictions of these interaction models for 3N systems are analyzed in
detail. It is shown that M-type models give, 
under certain conditions, a stronger binding of the 3N system than the 
original phase-equivalent model with nodeless wave functions. An 
analysis of the 3N system with the new versions of the Moscow NN 
potential describing also the higher even partial waves is presented. Large
deviations from conventional NN force models are found for the momentum 
distribution in the high momentum region. In particular, the Coulomb 
displacement energy $E_B(^3He)-E_B(^3H)$ - when  $E_B(^3H)$ is extrapolated 
to the experimental value displays a promising
agreement with experiment: $\Delta E_C\simeq $740 KeV. The validity and 
limits of two-body NN potentials in nuclei is discussed in the light of our 
analysis.

\end{abstract}

\section{Introduction}
\label{intro}
\par

It has been known for a long time that standard NN interaction
models based on one-meson exchange \cite{Bro76,Mac89,Had85,Bre60,Lac81}, 
meet significant difficulties in 
explaining  standard properties of few-nucleon systems, ordinary nuclei, as
well as nuclear matter \cite{Mac89,Had85}. In a number of cases, difficulties 
arise even in 
the well investigated area of low energies ($\sim 3-10$ MeV)
\cite{Wit91,Str88,Glo95}. 
In order to reach agreement between theory and experiment for
some key observables (binding energy, magnetic moments, etc.) 
3$N$-forces and meson-exchange currents (MEC) are very often introduced.
However, in practical applications, the values of  
cut-off masses are sometimes incompatible with those used in the initial 
two-body interaction model \cite{Mac89}.

These discrepancies have resulted in a revival of interest
in nonlocal nuclear force models (see e.g. \cite{Mac96,Mac93}). The latter 
give predictions closer to experiment than those of conventional local models
\cite{Mac96}. In fact, from the viewpoint of the more fundamental level of 
quantum chromodynamics (QCD) and the quark model, the
NN interaction must be strongly nonlocal at small distances 
\cite{Fae83,Fae93} 
due to quark exchanges between the nucleons (three-quark clusters) 
\cite{Fae83,Fae93,Obu91}.
A number of nonlocal models for the NN interaction,
either phenomenological \cite{Kis82,Sim84} or microscopic 
\cite{Fae83,Obu91,Oka83,Myh88,Val95,Lom85}, including both
meson and quark degrees of freedom have already been suggested. However,
most of them lead to rather complicated and bulky 
momentum-dependent NN interaction operators that can hardly be used 
for practical calculations in nuclear many-body systems. 
This explains the lack of {\em qualitative} studies of
the structure of few-nucleon systems using quark degrees of freedom.
Such studies would be extremely important for nuclear physics in general.  

Therefore, it is a rather urgent to develop a simple 
NN-interaction model which {\em simultaneously} takes into account 
both meson and quark degrees of freedom, and which 
is applicable to few- and many-nucleon systems. Having at  our
disposal a reasonably simple and realistic NN interaction model of the above
hybrid type, we can study its qualitative and
quantitative consequences in the many-body problem.

The purpose of this work is to do just that. At first, we develop a
simple nonlocal NN interaction model that is based on the 
quark structure of the nucleon.
We then study its predictions for the three-body bound state problem 
and compare these predictions with those of conventional force models.
An additional argument in favor of our approach is 
that quark degrees of freedom are not explicitly seen in the
energy region ($E^{NN}_{lab}\le $300 MeV). Only rather  
indirect signals of these hidden degrees of freedom can be studied. 
One such indirect quark effect
considered in the present work is the internal fermization of the nuclear 
wave functions \cite{Kuk94}. 
Such an "internal" fermization changes
the wave function of the system mainly, but not exclusively, 
at small distances and leads to the occurrence of additional internal 
nodes in the wave functions at about the same place where the repulsive 
core of conventional force models is situated.

From a formal point of view, the elimination of quark degrees of freedom 
results in additional orthogonality conditions for the NN channel in 
conformity with the classical Feshbach method for projecting onto mutually
orthogonal channels \cite{Fes62}. As a result, the orthogonality constraints 
lead to additional internal nodes in the wave functions \cite{Kuk86,Neu75}. 
Therefore, the height of the local repulsive core can be strongly reduced 
(see Sections III and IV). We emphasize that a large part of the repulsion 
in lowest partial waves is not due to the local core produced by 
vector meson ($\omega $ and $\rho $) exchange but comes from 
these additional orthogonality conditions. As will become clearer later on,
this leads to a substantial increase of the 
average nucleon kinetic and potential energy, which 
in turn has numerous consequences for the structure of 
the nucleus. Below we point out only the most important differences 
between conventional nuclear force models and M-type models.
\begin{itemize} 

\item[$\bullet$]
Because of the additional orthogonality 
constraints, the strength of the repulsive core, mainly due to $\omega$-meson 
exchange, can be considerably reduced down to values, dictated by 
$SU$(3) symmetry \cite{Mac89}, i.e. 
$g^2_{\omega NN}/4\pi \sim 5$. As a result, the 
attractive part of the potential is considerably deeper than
that of conventional force models. However, this does not cause any
discrepancy with experiment because the additional
orthogonality conditions lead to an effective reduction of the local
attraction at intermediate distances.

\item[$\bullet$]
Due to the sharp increase of kinetic energies in the NN channel
at small distances $r_{NN} < 1$ fm, the {\it relative} 
importance of nonlocal and energy-dependent
terms in the NN interaction operator coming from meson
retardation effects \cite{Mac89} or 6$q$-bags are strongly suppressed. As a 
result, all complicated and nonstatic
short-range nonlocalities originating from different sources are replaced 
by a very simple nonlocal {\em separable} potential. This nonlocal 
separable potential has the form of a projection operator and 
effectively describes the additional orthogonality conditions \cite{Kuk78}. 
This allows to drastically simplify the description of the NN interaction, 
especially at small internucleon distances.

\item[$\bullet$]
An important difference is also the {\em local} character of the
deep attractive potential which is universal for all partial waves. 
This universality can be interpreted as a consequence of pseudoscalar 
$\pi $, $\eta $ and scalar $\sigma$-exchanges between nucleons. Here,  
the scalar $\sigma $-exchange, gives a strong attraction 
(depth $\sim $500 MeV \cite{Fae83}) between nucleons at intermediate distances.
\end{itemize}

For sake of brevity, we shall call NN-force models, which
are characterized by (i) a strongly reduced local repulsive core 
(or even by the full absence of it), (ii) a strong attraction in the  
intermediate range and additional orthogonality conditions, as 
Moscow- (or M-) type models. For a substantiation of these models 
it is important to recall the following relatively new results.

In a number of recent studies \cite{Mic88,Bay87} it was established that a
limiting case of such models, which does not include any repulsive core at 
all, nearly coincides with the supersymmetric (SUSY) partner of the 
Reid-Soft-Core (RSC) potential. Thus, both types of interaction models,
conventional force models with a local repulsive core and M-type force 
models, are related by purely algebraic symmetry 
transformations. For a pedagogical discussion of this point 
see Ref.\cite{Kuk92}.

Recently Nakaishi-Maeda \cite{Nak95} has found that the phenomenological
separable NN-potential of Tabakin \cite{Tab68}, leading to internal nodes 
in the
radial deuteron and NN-scattering wave functions, is a simple
unitary-pole approximation (UPA) to the above Moscow NN-potential, i.e.
in the UPA-approach both models coincide with high accuracy.

These results, established by independent groups, are not accidental; rather
they are indicative of the quark dynamics underlying the Moscow NN-potential.
One of the main purposes of the present work is to explain this in greater 
detail and to study the qualitative behavior of M-type NN interaction models 
in the nuclear three-body problem.

This work is organized as follows. In Sect.2 we provide a  
substantiation of M-type models within the framework of the quark
model. Sect.3 presents a pedagogicial introduction to the main
physical ideas underlying the Moscow model. We study a simple scalar 
interaction and investigate its qualitative predictions for the 
three-nucleon bound state.
Sect.4 discusses the more complete model including tensor forces for 
positive parity partial waves.
In Sect.5 these more realistic models are applied to the 
three-nucleon bound-state problem. We emphasize the important
contribution of the tensor force to the 3N binding energy. The 
interference of central and tensor forces inherent in 
these models is quite different from conventional force models. 
The differences in predictions between traditional force models
and M-type models are pointed out. 
In the conclusion our main findings are summarized.

\section{Origin of NN interaction models 
with additional orthogonality conditions: M-type models }
\label{origin}

\subsection{Six-quark permutational symmetries and the NN interaction}
\label{permsym}

\par
Consider a two-nucleon system that is described in the 
framework of the nonrelativistic quark model \cite{Fae93,Obu91}. 
The quark Hamiltonian $H$ consists of the following components:
\begin{equation}
H=H_0+V_{OGE}+V_{conf}+V_{OME},
\end{equation}
where $H_0=\sum_{i=1}^6(m_i+ {p_i^2}/({2m_i}))$ is the kinetic energy
of the 6$q$-system. The interaction $V_{OGE}$ accounts for one-gluon
exchange, and the confinement interaction
$V_{conf}$ takes the standard form \cite{Fae83,Obu91}. In the 
one-meson exchange potential 
between quarks $V_{OME}$ we include, as usual, $\pi $ 
and $\sigma $-meson exchange potentials \cite{Obu91,Val95}. 

The totally antisymmetric 
(denoted by subscript A) wave function $\Psi _A(x_1,...,x_6)$ can be 
expanded into a sum of irreducible representations (IR) of the symmetry  
group $S_6$ \cite{Rac49,Ham64}. The sum extends over all allowed IR of 
the symmetry 
group $S_6$, characterized by Young schemes $[f]$, that lie in the outer 
product space $S_3 \otimes S_3$:
\begin{equation}
\Psi _A(x_1,...,x_6)=\frac {1}{\sqrt {n_f}}\sum _{[f]r}\Phi ([f]r;{\vec
r}_1,..., {\vec r}_6)\chi _{CST}([\tilde f]\tilde r),
\end{equation}
where the coordinates $x_1 \cdots x_6$ collectively 
stand for the position, and the spin, isospin, and color quantum numbers 
of the six-quarks. 
Here, $\Phi ([f] r; {\vec r}_1,..., {\vec r}_6)$ is the orbital part 
of the total six-quark wave function, and 
$\chi _{CST}([\tilde f]\tilde r)$ is its color-spin-isospin part.
The Young scheme of the space part of the six-quark wave function
is denoted by $[f] r$, where $r$ is the corresponding Yamanouchi symbol. 
The Young scheme $[\tilde f] {\tilde r}$ stands for the unique 
irreducible representation in spin-isospin-color space which is adjoint 
to $[f] r$.  For brevity we omit the Young symbols $[f]_C$, $[f]_S$,
$[f]_T$, relating separately to the color, spin and isospin parts of the
function  $\chi _{CST}$. Furthermore, $n_f$ is the dimension of the IR 
corresponding to the Young scheme $[f]$.

One can easily see that, according to the Littlewood theorem \cite{Ham64},
the allowed Young schemes, which describe the
permutational symmetry of six-quark orbital (X) wave functions $\Phi ([f]r)$,
are:
\begin{equation}
[3]_X\times [3]_X=[6]_X+[42]_X+[51]_X+[33]_X,
\end{equation}
where the first two terms correspond to even orbital angular 
momenta (with positive parity), and the last two terms to odd orbital 
angular momenta (with negative parity) of the relative NN wave function.
Thus, for even partial waves in the relative NN wave function,  the allowed 
space symmetries are $[6]_X$ or $[42]_X$, whereas for odd ones, they are 
$[51]_X$ or $[33]_X$.
Using the two-center shell model basis with distance $R$ between the two
centers,  one can show \cite{Sta87} that  all allowed six-quark wave
functions of the type $|S^3_+S^3_-[f] LST>$ approach the usual shell model
configurations $| s^mp^n [f] LST>(m+n=6)$ in the limit $R\to 0$. Thus,
the NN state with the totally  symmetric Young
scheme $[f]_X=[6]_X$ corresponds to the six-quark state $|s^6[6]_X[2^3]_{CS}
L=0, ST>$ while the NN state with mixed symmetry configuration
$[42]_X$ corresponds to a six-quark state with two excited $p$-shell quarks:
\[ | s^4p^2[42]_X[f]_{CS} L=0, ST>,\]
where $[f]_{CS}=[42],[321],[2^3],[31^3],[24^4]$ are the possible Young
schemes appearing in the Clebsch-Gordan decomposition of the internal product
$[2^3]_C[42]_S$.

In a series of works by many different authors 
\cite{Obu91,Oka83,Neu77,Obu88,Fae91}, it was shown 
that, with a quark Hamiltonian of the form of eq.(1), the excited six-quark 
configurations $|(0s)^5(1s)[42]_X>$ and $|(0s)^4(0p)^2[42]_X>$ 
which are compatible with $S$-wave relative NN motion, are  
admixed with large weights to the fully symmetric six-quark configuration 
$|(0s)^6[6]_X>$.  The color-magnetic term 
$\sim \b{\lambda}_i \cdot \b{\lambda}_j \, {\b{\sigma}}_i \cdot {\b{\sigma}}_j$
is mainly responsible for this. Thus, in lowest partial waves of the 
relative NN wave function,
there is a superposition of two different 6$q$-space symmetries: the fully 
symmetric $|(0s)^6[6]_X>$ and the mixed symmetric 
$|(0s)^4(0p)^2[42]_X>$  for $S$-waves, and similarly 
the $|(0s)^5(0p)[51]_X>$ and $| (0s)^3(0p)^3[33]_X>$  for 
$P$-waves.
It is important to remark that the above superpositions include the 
excited $p_{\frac {3}{2}} $ single-quark states. 

In recent work \cite{Obu91}, it was shown that six-quark components 
with different space symmetries play a very different role in the 
NN interaction. For example, the totally symmetric six-quark components 
$|(0s)^6[6]_X>$ are projected onto the NN cluster channel 
with rather {\em small weights}, whereas they have large projections 
onto the $\Delta \Delta $ and hidden-colour channels. In contrast, 
the  mixed symmetry components $| s^4p^2[42]_X>$  (in $S$-waves) 
have {\em large} projections onto both the cluster NN channel and 
the nucleon-isobar channels $N_1^*N_2^*$.

Thus, there is a natural separation of the complete six-quark wave function 
into two mutually orthogonal parts of a different physical nature:
\begin{equation}
\Psi_A =\Psi_{6q}+\Psi _{clust},
\end{equation}
where $\Psi _{6q}$ is the bag-like component, which can be constructed from
square integrable functions $\varphi _i$ corresponding to various 
six-quark bag states. These are the states with maximal space symmetry, i.e.
$s^6$ for relative $S$-waves and $s^5p$ for relative $P$-waves:
\begin{equation}
\Psi _{6q}=\sum_{i=1}^N C_i\varphi_i,
\end{equation}
where $N$ is the number of six-quark states.
For the cluster component we use the standard RGM-ansatz:
\begin{equation}
\Psi _{clust}({\b{\xi}}_1, {\b{\xi}}_2, {\bf R}) 
={\cal A}\left \{ \varphi _N ({\b{\xi}}_1) 
\varphi _N ({\b{\xi}}_2 )\tilde {\chi}({\bf R})\right \},
\end{equation}
where ${\cal A}$ is the antisymmetrizer with respect to the six quarks. 
Here, $\varphi _N ({\b{\xi}}_i)$ is the 
three-quark wave function of a single nucleon where the internal coordinates 
are collectively denoted by ${\b{\xi}}_i$; and 
$\tilde {\chi}({\bf R})$ is the relative motion wave function of two 
three-quark clusters.

\subsection{Two-phase model for the NN interaction}
\label{twophase}

\par
Our approach differs from the majority of hybrid $NN$ interaction models 
which are also based on a decomposition of the form of eq.(4). 
In contrast to, for example, the quark compound bag (QCB) model by 
Simonov [14], we require the mutual orthogonality of the components 
$\Psi _{6q}$ and $\Psi _{clust}$:
\begin{equation}
<\Psi _{6q}| \Psi _{clust}>=0.
\end{equation}
Moreover, we require that the cluster component be orthogonal to all
bag-like states $\varphi _i$ from which the component $\Psi _{6q}$ is
constructed:
\begin{equation}
<\varphi _i| \Psi _{clust}>=0, \qquad i=1,...N.
\end{equation}
When combined with eqs. (4)-(6), this leads to corresponding orthogonality
conditions for the RGM relative motion function $\tilde{\chi}(R)$:
\begin{equation}
<g_i| \tilde {\chi }>=0, \qquad i=1,...N,
\end{equation}
where
\begin{equation}
g_i({\bf R})=<\varphi _i({\b{\xi}}_1, {\b{\xi}}_2, {\bf R} )
| \varphi _N ({\b{\xi}}_1)\varphi _N({\b{\xi}}_2) >
\end{equation}
are projections of six-quark bag-like functions onto the $NN$-channel. To
emphasize the importance of orthogonality conditions in our approach we put
a $\sim $ sign over the relative motion wave function $\chi$.
We point out that a similar model with mutually orthogonal 
6q and NN channels in configuration space has been suggested by Lomon 
\cite{Lom85}. 
The model of Lomon is similar in spirit to our model but different in 
realization.

It is obvious that at intermediate and large distances, where the
cluster-like components dominate, the NN dynamics should be rather well
described by a meson-exchange (one-boson-exchange(OBE) and 
two-pion exchange(TPE)) model. Explicit quark and
gluon degrees of freedom are unimportant here. On the contrary, 
in the short-distance regime, with maximal overlap of the nucleon wave
functions, explicit quark-gluon degrees of freedom, described by the 
compound states $\varphi_i$  should play a decisive role.
In this way we arrive at what could be called a {\em duality principle} 
for the baryon-baryon interaction which can be formulated as follows. 
The total six-quark wave function 
naturally separates into two mutually orthogonal (i.e. nonoverlapping) 
components:

\begin{itemize}

\item[$\bullet$]
One component has a three-quark cluster structure where the baryons 
(including isobars) can be considered as separate entities, and where the 
dominant dynamical mechanism is meson exchange between the clusters, 
while the quark-gluon degrees of freedom are only of minor significance.

\item[$\bullet$]
The other component has a six-quark bag structure. Its dynamics is 
governed by explicit quark-gluon degrees of freedom. This component is 
sensitive to for example, the particular form of confinement, the value of 
the scalar quark condensate $<q\bar {q}>$, etc., and depends only weakly 
on the "external" meson dynamics.

\end{itemize}

Therefore, when constructing a proper NN interaction potential, the bag-like
components should be somehow excluded from the very beginning, because these
components are very hard to describe by any reasonable NN  potential.
For this component one should use a different formalism.
This is in complete analogy to the optical potential model in nuclear physics,
which is not at all applicable in situations, where the nucleon-nucleus 
scattering proceeds via isolated compound-states of the nucleus. In this 
analogy the optical nucleon-nucleus 
potential corresponds to the meson exchange dynamics between nucleons
while the nucleon-nucleus compound states correspond to the six-quark bags
in NN scattering.

Projecting the total six-quark Schr\"odinger equation
$$H\Psi =E\Psi $$
onto the six-quark states $\varphi _i$ and onto the NN cluster channel 
we obtain, using
the orthogonality conditions (8), the following set of coupled equations:
$$\left \{
\begin{array}{rcll}
\sum _j(\hat {H}_{ij}-E\delta _{ij})C_j  + <f_i| \tilde {\chi }>& = & 0, &
\qquad \qquad \qquad \qquad  (11)\\
\sum _jC_jf_j+(\hat {\cal H}^{RGM}-E\hat {\cal N})\tilde {\chi }&= &0. &
\qquad \qquad \qquad \qquad (12) \\
\end{array}\right \}$$
Here, we have introduced the following abbreviations
\setcounter{equation}{12}
\begin{equation}
\hat {H}_{ij}=<\varphi _i| H| \varphi _i>,
\end{equation}
\begin{equation}
f_i({\bf R})=<\varphi _i| H| \varphi _N\varphi_N>.
\end{equation}
$\hat {\cal H}^{RGM}$ and $\hat {\cal N}$ are integral operators with
kernels:
\begin{equation}
{\cal H}^{RGM}({\bf R},{\bf R}')=<\varphi _N\varphi _N| H{\cal A}| \varphi
_N\varphi _N>,
\end{equation}
\begin{equation}
{\cal N}({\bf R},{\bf R}')=<\varphi _N\varphi _N| {\cal A}| \varphi
_N\varphi _N>.
\end{equation}

Solving the linear algebraic equations (11) with respect to the 
coefficients $C_i$ and 
substituting its solution
\begin{equation}
C_i=\sum _j(E-\hat {H})^{-1}_{ij}<f_i| \tilde {\chi }>
\end{equation}
into eq. (12), we find the equation of motion for the NN relative 
wave function $\tilde {\chi }$:
\begin{equation}
\left (\hat{\cal H}^{RGM}-E\hat{\cal N}+\sum _{ij} | f_i>(E-\hat
{H})^{-1}_{ij}<f_i| \right )\tilde {\chi }=0
\end{equation}
This equation must be solved together with the additional orthogonality
conditions (9). After that one can calculate the coefficients $C_i$ via eq.
(17) and hence the total wave function $\Psi $.  
Eq. (18) constitutes our two-phase model for the NN interaction.

Due to the appearance of the norm kernel ${\cal N}$, 
the RGM relative motion wave function 
$\tilde {\chi }$ in eq. (18) cannot yet be interpreted as the 
probability  amplitude for finding the nucleons at a relative distance $R$. 
In order to obtain a NN relative motion wave function that can be interpreted
in the usual sense,  we define a renormalized function $
\hat{\tilde  {\chi }}={\cal N}^{1/2} \tilde {\chi} $
and a corresponding effective nucleon-nucleon interaction:
\begin{equation}
(T_R+V_{eff}-E)\hat{\tilde {\chi }} = 0,
\end{equation}
where
\begin{equation}
V_{eff} = {\cal N}^{1/2}\hat{\cal H}^{RGM}{\cal N}^{1/2}
-T_R+\sum_{ij} | \hat{f}_i>(E-\hat {H})^{-1}_{ij}<\hat {f}_i|,
\end{equation}
$$\hat {f}_i=\hat {\cal N}^{1/2}f_i.$$
The effect of the overlap kernel ${\cal N}({\bf R},{\bf R}')$ for the NN 
system can be reasonably well approximated by the expression \cite{Obu91}:
\begin{equation}
{\cal N}({\bf R},{\bf R}') \simeq \frac {1}{10}\delta ({\bf R}-{\bf R}'),
\end{equation}
which differs from the usual local NN potential just by a 
combinatorial factor.
Thus, the effective interaction can be cast into the form:
\begin{equation}
V_{eff} = V_{dir}+V_{ex}+V_{NqN},
\end{equation}
where $V_{dir}$ is the direct (folding) potential, including one-meson
exchanges between three-quark clusters only, i.e. the one-meson-exchange
nucleon-nucleon potential. Because the one-gluon exchange operator is 
diagonal with respect to quark permutations between nucleons it does not
contribute to $V_{dir}$.
$V_{ex}$ is the nonlocal short-range exchange potential, 
and $V_{NqN}$ is an
effective potential due to the coupling of NN- and $6q$-channels:
\begin{equation}
V_{NqN}=10\sum_{ij}| f_i>(E-\hat {H})^{-1}_{ij}<f_j| .
\end{equation}

The effective two-nucleon Schr\"oderinger equation for the orthogonalized 
and renormalized relative motion
wave function $\tilde {\chi }(R)$:
\begin{equation}
\left \{
\begin{array}{rcll}
(T_R+V_{dir}+V_{ex}+V_{NqN}-E)\hat {\tilde \chi } &=& 0 &\\
<g_i| \hat {\tilde \chi }>&=&0,& i=1,...N\\
\end{array}\right.
\end{equation}
provides the basis for the justification of NN-interaction models of 
Moscow type.

The main point is that the solution of  eq.(24) is defined
only in the {\em subspace orthogonal} to the functions $g_i$. 
Due to this orthogonality requirement, we 
inevitably introduce nodes in the NN scattering wave functions 
at small distances \cite{Kuk86,Neu75,Neu77,Obu88,Fae91,Sai69}. 
The node positions are very stable as the NN scattering
energy $E$ is increased \cite{Obu91,Kuk86}. This behavior of scattering wave 
functions can be quite naturally formulated in the language of interaction 
potentials that have a very deep attractive well ($\sim $1 GeV) \cite{Neu75}
with additional deep-lying bound states. In fact, compared to the deep attractive
well with $|V_0| \approx 1 $ GeV, the relative scattering energy $E$ is 
negligible. Therefore, the position of the innermost node $r_{node}$ 
is basically determined by the well depth $V_0$:
$$\sqrt{m_N\,(E-V_0)\, }r_{node} \approx \sqrt{m_N\,|V_0|\, }r_{node} \approx \pi. $$  
Thus, the inner node of the scattering wave 
function is almost stationary for center of mass energies up to 
$\sim $2 GeV in the laboratory frame in nonrelativistic kinematics. 
Inserting the numerical values for the nucleon mass, and $V_0$ we find
that the stationary node is located at $R_n\simeq $0.6 fm in the case 
of relative $S$-waves ($R_n\simeq $0.9 fm in the case of $P$-waves).
We point out that this node plays the role of the repulsive core in 
the NN interaction.

Because the nodal solutions of the basic equation (24) with orthogonality 
conditions correspond to very large kinetic energies, 
the short-range nonlocal potential $V_{ex}$, which includes the quark
exchange diagrams, plays only a minor role in the subspace of nodal 
wave functions. This is in analogy to the nuclear Orthogonality Condition 
Model (OCM) of Saito \cite{Sai69}. Therefore, $V_{ex}$ can be omitted from 
equation (24) in good approximation.

Finally, we take the last step in the substantiation of the
model. In absence of the  vector ($\rho $ and $\omega $) meson contribution
in the total Hamiltonian (1), the meson-exchange potential $V_{dir}$ in
eq. (24) corresponds to $\pi $ and $\sigma $-meson exchange. 
Therefore, it is strongly attractive with a depth of the central
potential of about $\sim $500 MeV. Together with the nonlocal 
attractive (for $E<E_1$)
part $V_{NqN}$, it results in an {\em effective} deep ($\sim $1
GeV) attractive potential, which was found in our initial
attempts \cite{Neu75}, and has since been justified 
\cite{Obu91,Kuk86,Kuk92,Glo92} within the 
framework of the quark model.
In fact, we have found out only quite recently \cite{Kuk97} that
the combination of the local attractive $\sigma $-meson 
exchange potential (with a depth of about 500 MeV) and the 
nonlocal attraction $V_{NqN}$, representing the coupling of the NN and 
6q-bag-like channels, is 
exactly phase-shift equivalent to the deep ($\sim $1 GeV) effective {\em 
attractive} potential of  M-type.

The Moscow potential is the simplest local model, which ensures the
orthogonality of its solutions to a function $\varphi_0$ that approximately 
describes the 
projection of the fully symmetric $(0s)^6$ six-quark state onto 
the NN-channel (in $S$-waves).
We emphasize that the occurrence of unobservable deep-lying bound states 
in the equation (24) poses no problem, because the equation for the NN channel
wave function is considered in the appropriate subspace of nodal wave 
functions, which is {\em orthogonal} to the nodeless bound states.
Additional extra deeply-bound states arise 
only in the solution of the appropriate Schr\"odinger equation in 
the {\em complete} space.

In order to understand the reasons for the occurrence of deep-lying bound
states, it is very instructive to recollect \cite{Obu91} that, the weight of 
the NN-configuration (with {\em unexcited} nucleons) in symmetric bag-like
six-quark states $| s^6[6]_X[2^3]_{CS} L=0, ST>$ is very small; whereas
the weight of the $\Delta \Delta $ channel is appreciably larger and the
main contribution comes from hidden colour $CC$ channels, whose wave functions
are localized (because of confinement) at small distances \cite{Mat77}. As a
result of the strong coupling of the initial NN channel and the 
bag-like six-quark channels, the probability for finding an 
NN configuration with unexcited nucleons vanishes almost 
completely at small distances\footnote{A similar
conclusion concerning the suppression of the NN-channel in favor of the
$\Delta \Delta $- and other isobar channels can possibly be derived
in a formalism based on meson Lagrangians \cite{Sau87} without any reference 
to the quark model.}.
Thus, at very small distances, where the extra deeply-bound states are
localized, there is basically no nucleon-nucleon configuration.
In this regime, the 2$N$-system  can be viewed as though "being dissolved" in 
a quark "soup". Remembering the Cheshire Cat smile, it is appropriate to name 
these deep-lying bound states: "Cheshire-Cat Bound States (CCBS)".

In applications of the model to many-nucleon systems, it should be
kept in mind that the Moscow potential describes in an effective way the
single-channel system, i.e. the 2N system with {\em unexcited} nucleons, 
while the real 2N-system should be
described by the above two-component model. Thus the description of
three-nucleon systems given by the Moscow {\em potential } model must be
supplemented with an additional 6$q$+N contribution which is expected to be
strongly {\em attractive}. Before we discuss the Moscow NN potential
in greater detail, we explain the main ideas of our approach within the 
framework of a simple toy model.

\section { A simple M-type model for the NN-interaction} 
\label{simple}

\subsection{General considerations}
\label{general}

In this section we compare two alternative models for the NN-interaction.
For clarity, we use the simplest possible example, i.e.  
two spinless nucleons interacting via a scalar potential $V(r)$:
\begin{equation}
V(r)=g_R\, V_R(r)+g_A\, V_A(r).
\end{equation}
The potential includes a short-range repulsive core $g_R\, V_R(r)$, 
and a long-range attraction 
$g_A\, V_A(r) $. The coupling constants $g_R$ and $g_A$ 
are chosen such that the system has 
a single bound state with energy $\varepsilon _0$ ("deuteron") and 
scattering phase shifts $\delta _l (E) $, $l$=0,1,....

Let us try to find a modified potential $V'(r)=
{g'}_R\, V_R(r)+g_A\, V_A(r)$ with a reduced
strength of the repulsive core, but simultaneously introduce an 
additional constraint that ensures the orthogonality of the solution of 
the Schr\"odinger equation
to some function $\varphi _0$ localized in the core area:
$$\left \{
\begin{array}{rcll}
(T_r+V'(r)-\varepsilon ) \tilde \psi (r) = 0 \qquad  \qquad \qquad  \qquad
(26a)\\
\langle \varphi _0| \tilde \psi (r) \rangle = 0 \qquad \qquad  \qquad
\qquad (26b)
\end{array}\right.$$
The physical meaning of the additional orthogonality constraint has already 
been 
discussed and partly been elucidated above. Here,  we consider 
this orthogonality condition as a formal constraint that the solution 
has to satisfy.

Equation (26a) with the additional condition (26b) can be rewritten in
the form of a single equation with the nonlocal interaction (see \cite{Sai69}):
$$(T_r+V'(r)-\varepsilon )| \tilde {\psi }(r)\rangle =| \varphi
_0\rangle \langle \varphi _0| T_r + V'(r)| \tilde \psi (r)
\rangle. \eqno (27) $$
The solutions of this equation are, as one can easily see, orthogonal 
to $\varphi_0$ for any $\varepsilon \ne 0$. Only for $\varepsilon
$=0, the required orthogonality is not guaranteed and Eq.(27) 
gives an incorrect behavior for the corresponding off-shell $t$-matrix.

One can apply a more convenient approach, known 
as the method of orthogonalizing pseudo-potentials (OPP) \cite{Kuk78}.
In this approach, the additional orthogonality condition (26b) is taken 
into account via the projection operator $\Gamma = | \varphi _0\rangle
\langle \varphi _0| $ which projects onto the "forbidden" subspace
with a large coupling constant $\mu$:
$$ \left (T_r+V'(r)+\mu | \varphi _0\rangle \langle \varphi _0|
-\varepsilon \right )| \tilde \psi _\mu (r)\rangle_{\mu \to \infty}=0. 
\eqno(28)$$
In our example, the forbidden subspace is just a one-dimensional subspace 
spanned by the vector $\varphi _0$. 
In the limit
$\mu \to \infty $, the solution of equation (28) 
$\tilde \psi _\mu (r)$ have been shown \cite{Kuk78} to be rigorously 
orthogonal to $\varphi _0$, i.e.: 
$$\lim _{\mu \to \infty }\langle \varphi _0| \tilde \psi _\mu \rangle =0.$$
For $\varepsilon \ne 0$ the solution $\tilde \psi _\mu $ coincides with the
solution of the  Saito equation (27).
The projection operator $\Gamma = | \varphi _0\rangle \langle \varphi _0|
$ defines a "forbidden" subspace ${\cal H}_\Gamma $ of the full
two-particle Hilbert space ${\cal H}$ which can be decomposed as 
$${\cal H}_\Gamma \bigoplus {\cal H}_Q= {\cal H} $$
where ${\cal H}_Q$ is the orthogonal complement to ${\cal H}_\Gamma $.

Thus, we consider the problem of finding a modified potential $V'(r)$ that 
is phase-shift equivalent to the initial potential $V(r)$ of eq.(25) and 
which has
the bound state at the same energy $\varepsilon _0$ in the {\em subspace}
${\cal H}_Q$ orthogonal to $\varphi _0$. 
For an arbitrary $\varphi _0$, 
rigorous solutions of such an inverse scattering problem are, 
to our knowledge, not available. 
Therefore, we search for approximate solutions. In order to further elucidate 
the relation of M-type models and conventional NN force models, we are
interested in such solutions that contain a repulsive core that is weaker 
than that in the initial potential $V(r)$. Although the modified 
potential $V'(r)$
becomes deeper when the core is weakened, the introduction of the additional
orthogonality constraint of eq. (26b) renders the
Hamiltonian {\em effectively} weaker. Due to this compensation, both the
phase shifts and the bound state energies remain the same, albeit with
some error.

On the other hand, if we use the OPP-method \cite{Kuk78,Sai69} the  modified
pseudo-potential  has the form:
$$\tilde {V}'(r)\mathrel {\mathop { = } \limits _{\mu \to \infty }}
V'(r)+\mu
| \varphi _0\rangle \langle \varphi _0| \eqno (29) $$
Using this method, it is possible to work {\em in the total} space instead of 
only in the subspace. This is one of the practical advantages of the 
OPP-method. Thus, in the OPP approach, a part of the {\em local} short-range 
repulsion 
is replaced by a separable repulsive potential with an infinite coupling
constant $\mu$ (29). If $\varphi _0$ is a bound-state solution of the   
potential $V'(r)$, the potentials $V'$ and $\tilde {V}'$ are completely
phase-shift equivalent. For this case, the solution of the inverse problem,
i.e. the transition to a phase-shift equivalent potential with a local 
repulsive core {\em without} a deep-lying bound state, is well known. 
This is a supersymmetrical (SUSY)
transformation \cite{Mic88,Bay87}. However, when extra bound 
states (number $n$) are eliminated, 
a SUSY-transformation always introduces a repulsive core of centrifugal type, 
i.e. $v_s (r)\mathrel {\mathop {\sim } \limits _{r\to 0}} n (n+1)/r^2$, 
instead of the usual Yukawa-type core in local OBEP models. 
Furthermore, the invariance of phase shifts under a SUSY
transformation holds only for a {\em given} partial wave.

Here, we want to consider the more general case of a repulsive core of 
intermediate strength. Furthermore, we want to preserve the invariance of 
the phase shifts in several not just in one 
partial waves. Most importantly, we do not want to introduce any 
superfluous bound states into our model from the very beginning. 
Therefore, we refrain from using the rigorous results of SUSY-transformations 
and we look for a potential that is close to the phase-equivalent potential 
in the subspace ${\cal H}_Q$ orthogonal to a given bound state.

\subsection{The Malfliet-Tjon potential}
\label{malfliet}

\par
To be specific let us consider a particular example. 
As the initial potential we take the Malfliet-Tjon potential 
MT-V \cite{Mal69}. The MT-V potential is a simple scalar model for
the NN interaction, which yields an average deuteron binding energy of 
$-0.4136$ MeV and describes the {\em average} $^3S_1$ and $^1S_0$
NN-phase shifts reasonably well up to energies of about 300 MeV:
$$V_{MT-V}(r)=g_R\frac {\exp (-\mu_R \,r )}{\mu_R \, r}+g_A\frac {\exp
(-\mu_A \, r)}{\mu_A \, r},\eqno (30)$$
where the parameters values are
$$g_R=1458.05 \, {\rm MeV}, \; g_A=-578.09 \, {\rm MeV}, \; 
\mu _R=3.11 \,  {\rm fm}, \;
\mu_A = 1.55 \, {\rm fm}. \eqno (30a)$$
Next, we introduce  an orthogonality condition  to the function $\varphi
_0(r) $ of Gaussian form:
$$\varphi _0(r)=N\exp \left (-\frac {r^2}{2r^2_0} \right ). \eqno (31)$$
This (0$s$) harmonic oscillator (h.o.) function 
approximates quite well the projection
of the six-quark bag  $| s^6[6]\rangle $ 
onto the NN  channel \cite{Obu91,Neu75,Kuk92,Mat77}. For various values 
of the h.o. radius
$r_0$ we now determine the coupling constants $g'_R$ and $g'_A$ of
the modified potential:
$$V'(r)=g'_R\frac {\exp (-\mu_Rr)}{\mu_Rr}+g'_A\frac {\exp
(-\mu_Ar)}{\mu_Ar} \eqno (32)$$
{\em acting in the subspace} ${\cal H}_Q$, which ensure the best fit to 
the $S$-wave
phase shift for the {\em initial } potential MT-V (30), acting in the
{\em full} space ${\cal H}$. For the sake of simplicity we do not change 
the inverse radii of the core, $\mu _R$, and of the attractive potential
$\mu _A$.

The results of this fit for different values of radius $r_0$ are
presented in Table 1. It can be seen that a fit is possible for 
$r_0$ lying in the interval 0.15$\div $0.5 fm. Thus, in the 
{\em subspace}
${\cal H}_Q$ orthogonal to $\varphi _0(r)$, the partial $S$-wave phase shifts
for the  potential $V'(r)$ of eq.(32) reproduce fairly well the corresponding
phase shifts of the initial potential (30) in the energy range $0\div 300$
MeV (see the $\chi ^2$ values  in Table 1). We do not present here the
respective figures because the difference between the phase shifts of the 
initial potential MT-V and those of $V'(r)$ is almost indistinguishable 
in the graph.

It should be noted that the local core disappears already at $r_0$=0.2 fm. A
further increase in $r_0$ results in an increase of the radius $r_{node}$ of
the internal node in the wave function and of the
amplitude of internal loop specified by the ratio $\beta $ of absolute
values of the wave function in the first and in second maxima (see
the fifth column in the Table 1). After such a modification of the  potential, 
the bound state energy
$\varepsilon _0$ ({\em in the full space}) varies from $-2.51$ up to $-600$
MeV. However, if we require the orthogonality to $\varphi _0$, i.e. if we are
operating {\em in the orthogonal subspace}, the bound-state energy 
$\tilde {\varepsilon }_0$  in Table 1 remains almost constant that is 
of the order of 1 MeV. 

Thus, we obtain a series of approximately 
spectral-equivalent Hamiltonians\footnote {In contrast to trivially
spectral-equivalent Hamiltonians that are obtained as a result of the
unitary transformation $H'=U^{-1}HU$ and which have been investigated 
in detail in the literature \cite{Sau75}, our transformed potentials 
do not include
any velocity dependence. Most importantly however, they have a different 
physical meaning.}.
The best (i.e.the smallest)value of $\chi ^2$ is obtained for 
$r_0$=0.4 fm when there are two bound states in the potential $V'(r)$:
the first one is deep-lying with the energy 
$\varepsilon _0=-528$ MeV ÿ(see
the third row in Table~1) and the second one is close to the NN-threshold
at the energy $\varepsilon _1=-0.4$ MeV, the eigenfunction of the deep
bound state being very close to the h.o. function $\varphi _0(r)$. In other
words, we arrive again at a deep attractive potential {\em
without} any local repulsive core but with an eigenstate projector, that is,
to a construction that is very similar to the Moscow potential model.
Thus, there are two extreme cases: (i) the deep purely attractive potential 
with the "extra" deep-lying state at $\varepsilon _0=-528$ MeV corresponding to
$r_0$=0.4 fm,
%We point out that the h.o. wave function 
%$\varphi _0(r)$ with this value $r_0$ is very close to the {\em exact} 
%wave function of the deeply-bound state in this potential.
and (ii) the initial MT-V potential which does not contain the "extra" 
deep-lying bound state state at $\epsilon_0=-528 $ MeV but has a local 
repulsive core instead.

These two phase shift equivalent potentials are approximate supersymmetric
partners. All other cases in Table~1 can be considered as intermediate 
variants, lying between these extremes\footnote{Such an intermediate 
model may be relevant for modelling the physical $\omega$-meson exchange
interaction with a reduced $\omega$NN coupling strength.}.   
By changing the
"projector radius" $r_0$ from zero to 0.4 fm, we perform a continuous
transformation from one extreme model to the other. In doing so, the position
$r_{node}$ of internal node of wave function changes and the amplitude of
the internal loop of wave function increases. This can be seen in Fig.~1,
which shows the scattering wave functions at 1 MeV for three potentials:
the initial local MT-V (dashed line 1), the intermediate nonlocal 
potential with $r_0$=0.2 fm (solid curve 2) and the 
potential having an approximate eigen state projector with $r_0$=0.4 fm 
(dotted curve 3).

We emphasize again that, the second extreme case with $r_0$=0.4 fm involves  
an almost exact eigen state projector, i.e. for the given values of the 
coupling constants $g'_R$ and
$g'_A$ (see Table~1) the modified potential (32) leads to a deep-lying
bound state with with energy $\varepsilon _0\simeq -528$ MeV.
The corresponding eigen function is very close to 
$\varphi _0(r)$ at $r_0$=0.4 fm. Because the Hamiltonian is
Hermitian, all scattering wave functions, and also the eigenfunction
of the second (near-threshold) bound state at $\varepsilon _1=-0.41$ MeV are
automatically orthogonal to the wave function of the deep-lying level.
Hence, there is no need for any additional orthogonality conditions.

As a result, the supersymmetric partners are {\em local} potentials,
while all phase-equivalent intermediate cases correspond to  {\em nonlocal}
interactions in the full space, or alternatively, to  {\em local}
interactions {\em in the subspace} ${\cal H}_Q$.
However, according to our intention, we would like to obtain a good
description not only in partial $S$-waves, but also in  
$D$-waves and possibly
in other even partial waves. In a realistic hybrid model of the baryon - baryon
interaction, a quark bag is formed only in the lowest partial waves.
Therefore, the additional orthogonality condition must be included only in
lowest partial waves ($S$- and $P$-), whereas {\em the same Hamiltonian} 
without any additional orthogonality constraints is expected to correctly 
describe the phase shifts in higher partial waves.

Thus, we attempt to find a modified potential (32) acting in the 
subspace ${\cal H}_Q$ that {\em simultaneously} describes the 
$S$- and $D$-wave phase shifts for the initial MT-V potential. 
This is possible with acceptable accuracy for various values of 
$r_0$. Table 2 presents several variants of the modified potential (32) 
with different values $r_0$ that are fitted to the $S$- and 
$D$-wave phase shifts under the condition that these are orthogonal to the
$S-$ wave function $\varphi _0$. Because we intend to use the above 
potentials in three-body calculations, we present here the versions for which 
the 
"deuteron" binding energies are practically identical ($\equiv $0.413 MeV). 
The results of three-body calculations with these potentials is also given 
in the Table~2 and are discussed below.

Figs.~2 and 3 show a comparison of the $S$- and $D$-wave phase shifts in the
energy range 0-400 MeV for three potentials: (i) the initial local MT-V
model, (ii) a modified nonlocal potential $V'(r)$ with $r_0$=0.2475 fm 
(i.e. a potential without a repulsive core that is close to the
SUSY-partner of the MT-V),  and  (iii) a modified nonlocal potential $V'(r)$ 
with $r_0$=0.15 fm (an intermediate variant).
It is clearly seen that the potential with the strongly reduced repulsive core
acting in the subspace ${\cal H}_Q$ provides a satisfactory description of 
both $S$- and $D$-wave phase shifts of the initial MT-V potential 
acting in the full space ${\cal H}$.

In this way, our simple toy-model shows that the introduction of an
additional condition which ensures orthogonality to some localized state
$\varphi _0$ enables us to perform a {\em continuous transition} between 
the local core model and the model with "extra" bound states but no core.
Both alternatives describe the interaction between composite nucleons equally 
well. The extreme models happen to be SUSY partners.
In between, one has a series of almost phase shift equivalent potentials, 
differing by the core strength and by the spatial extension $(r_0$) of 
the state $\varphi _0$.

\subsection{Consequences for three-nucleon bound states}
\label{consequences}

\par
Next we discuss, how the properties of the 3N-system change when 
going from one NN-force model to the other. In table 2 
we compare the 3N predictions of the initial MT-V model (30, 30a) with those 
of the different variants of the orthogonalized model.
Table~2 displays bound-state energies $E_3$, kinetic energies $T_3$,
and also the minimum and secondary maximum position of 
the $^3$H charge form factor together with its value $F_{max}$ 
in the secondary maximum for different variants of the potential (32) and
different functions (i.e. different values of $r_0$) $\varphi _0(r)$. 

One can see that an increase in the "projector radius" $r_0$ from 0 
up to 0.25 fm leads to {\em an increase} of the three-body binding energy from
8.25 up to 8.99 MeV. As $r_0$ increases further up to 0.4 fm the binding
energy begins to decrease again. These results contradict the results of
Nakaishi-Maeda \cite{Nak86}, who showed, that the replacement of the local 
core by
the orthogonality condition leads always to a {\em reduction} of the
three-particle bound-state energy. According to \cite{Nak86}, the "stronger" 
the imposed orthogonality condition, i.e. the larger the amplitude of 
the inner loop in the radial wave functions, the more pronounced the 
reduction of the binding energy. At first sight, our results seem to 
contradict the 3N-calculations with the Moscow potential, undertaken by 
several groups (see e.g. \cite{Hah86,Kam97,Lee94}). 
These calculations showed that the 
replacement of the local
repulsive core NN-potential (for example, the Reid potential) by the Moscow
$NN$-potential leads to a {\em reduction} of the binding energy from 7 down to 
4.5 MeV \cite{Hah86}.

The difference between our results and the Nakaishi-Maeda findings is 
explained as follows. The transition from the local core to the 
orthogonality condition causes a sharp increase in the kinetic energy of the 
3N system (see the fifth column in Table 2). This in turn  leads to an 
increase of the contributions of higher partial waves. 
The main role is played by an attraction in $D$-waves - 
see the seventh column in Table. 2. It should be emphasized that the MT-V 
potential is central and therefore does not mix three-body states
with different values of the total angular momentum $L$. 
We consider here the mixing {\em within} a state of given total angular 
momentum $L$ (e.g. $L$=0). By higher
partial waves, we mean states with $\lambda =l=2,4,...$, where $\lambda $
and $l$ are angular momenta associated with the Jacobi coordinates  $r$ and
$\rho $, respectively.

The change in the 3N binding energy thus depends on the magnitude and sign of
the $D$-wave interaction: if this interaction is strongly attractive, the 
binding energy is increased, if it is repulsive the binding energy is 
decreased. To verify this statement, we have carried out 3N calculations with 
strong repulsion in $D$-wave interactions, while keeping the $S$-wave 
interaction fixed. The results are presented in the eighth column of Table 2 
($E^{rep}_3$).
With increasing $r_0$, i.e. with a strengthening of the orthogonality
constraints, the contribution of the internal $D$-wave grows and the binding
energy $E^{rep}_3$ with repulsive $D$-wave interaction monotonically
decreases.

The calculations of Nakaishi-Maeda \cite{Nak86} and Hahn et al. 
\cite{Hah86} were based on
the Faddeev equations and ignored usually small contribution of
higher partial waves. Under these restrictions, our calculation also shows
that the orthogonality condition always leads to a stronger effective
repulsion than the original local repulsive core model.\footnote{This is 
probably due to the fact that the radius of the separable repulsive core 
in the pseudo-potential $\mu | \varphi _0> 
<\varphi _0| $ (for $\mu \to \infty $), is much greater than the radius of 
the original local repulsive core.}
As a result, without the higher partial waves the binding of the 
3N system becomes weaker as one goes from the local
repulsive core model to an M-type model. If we include the 
contribution of higher partial waves in pair subsystems (for a fixed 
value of the total orbital angular momentum $L$), the 3N binding energy 
changes in the way described above. An appreciable increase of the 
contribution of the higher partial waves in pair subsystems 
due to the increase of the internal kinetic energy is the distinctive 
feature of M-type models.
A more detailed discussion of this property is given below when we 
consider a more realistic model.

\section{The Moscow potential model}
\label{Moscow}

\subsection{One-channel model}
\label{onechannel}

\par
Early attempts to construct an NN-potential with an additional "forbidden"
state, i.e. a model of M-type, were undertaken already in the middle of the
seventies \cite{Neu75}. However, in these early works only central 
potentials and only even  partial waves were considered. We point out 
that these early attempts were undertaken yet
{\em before} the color degree of freedom quark was fully established. 
But for colourless quarks
(fermions), the lowest (as it then seemed) six-quark configuration $|
(0s)^6[6]_X L=0,ST>$  is strictly forbidden by the Pauli principle.
Therefore, these NN potentials were interpreted in complete analogy with 
the effective potentials between nuclear clusters (e.g. the $\alpha -\alpha $ 
potential), in other words as potentials with
bound states forbidden by the Pauli principle.

The first version of a realistic Moscow potential was suggested in 1983 
\cite{Kuk83} and further details can be found in Ref. \cite{Kuk85}. 
This version 
describes the NN interaction only in the $^1S_0$ and $^3S_1-^3D_1$ channels.
Later on the description was extended to the lowest odd parity partial waves (
$^3P_0$, $^3P_1$, $^3P_2-^3F_2$) \cite{Pom86}. A new and improved version of
the Moscow potential (version 86) was published in \cite{Kuk86a}. 
This version still
had  a node in the deuteron  $D$-wave for which there is no evidence in 
microscopic quark models \cite{Fae83,Obu91,Oka83}. Finally, in 1990,
we have changed the truncation of the tensor OPE potential 
\cite{Kuk92}, which
eliminated the $D$-wave node (version 90).

The main difference between the Moscow potential and standard models for the 
NN interaction is the absence of a local repulsive core at small distances.
Instead of the core, the interaction is described by a deep attractive
potential $V_0\exp(-\eta r^2)$ 
plus an appropriate condition of orthogonality to an extra deeply bound
state. Owing to the deep-lying bound state, the physical NN wave functions 
have a node at small distances. The position of this node at around 
$\sim 0.6$ fm is almost stationary with increasing energy. In our approach, 
this stationary node plays the role of the repulsive core 
in standard models, and  provides the
correct behavior of the NN scattering phase shifts. The physical meaning 
of the "extra" bound states in the lowest partial waves has been repeatedly 
discussed in the literature \cite{Kuk86,Kuk92,Glo92}. They either simulate 
six-quark bag
compound states, for example,  $|s^6[6]_X\rangle $ in $S$-waves, 
and $| s^5p[51]_X \rangle $ in relative $P$-waves, 
which cannot be adequately described in terms of nucleon 
degrees of freedom only \cite{Obu91,Sim84,Myh88,Kuk86}; or they 
describe bound states in the 
nucleon plus Roper resonance NN$^*$(1440) channel \cite{Glo92}. 
We recall that the 
interaction in the NN$^*$(1440) channel must be very similar to the original 
NN interaction because the Roper resonance has the same quantum 
numbers as the nucleon.

Because we are mainly interested in the predictions of Moscow-type
NN-potentials for 3N systems, a more complete version of such 
potentials including even partial waves with $l\le $4 is necessary. Here, we 
present a few new versions for {\em even} partial waves. The form of the 
interaction for odd partial waves (for the version 86) is discussed 
in \cite{Kuk86a}. An updated version of the Moscow potential for odd
partial waves will be considered in a future publication.
In accordance with general expectations, six-quark bags
appear mainly in the lowest $S$ and $P$ partial waves 
(for $E_{Lab} \le 800$ MeV). Therefore, the extra bound states describing the
projections of these six-quark states onto the NN channel can occur only in
these lowest partial waves.

In the higher partial waves there are no extra bound states and 
correspondingly no nodes in the wave function. However, an interaction 
that has in the lowest partial wave only an attractive local potential 
and a repulsive projector multiplied by a large positive constant, 
is too strong for 
higher partial waves. In order to take into account the short-range repulsion 
generated by $\omega $-meson exchange, and in order to 
retain the universal form of the interaction in all partial waves, we add a 
repulsive core in the same separable form with a Gaussian form factor, 
as in the lowest partial waves, but with a {\em finite} positive coupling 
constant. We will then arrive at some {\em intermediate} model of the type 
discussed in the previous section. 

The final Moscow potential consists of two parts. First, 
a local $l$-independent part, which explicitly includes the one-pion-exchange 
(OPE) potential and a deep attractive well $V_0\exp(-\eta r^2)$ 
that depends on the spin and 
parity of the NN system. Second, a state-dependent separable repulsive core 
with Gaussian form factor, which provides the correct  $D$-wave  
phase shifts. This form of the potential is not only universal, but it is also
convenient for three-body variational calculations, because it is possible to
calculate all of the matrix elements of the Hamiltonian analytically
\cite{Kuk75,Kuk77,Kam89}. The separable core providing the short-range 
repulsion depends 
on $l$ and $J$. Therefore, it is not necessary to explicitly  include 
a spin-orbit interaction for even partial waves.

The OPE potential is truncated in a suitable way at small distances 
(see below). The tensor interaction which couples partial waves with angular 
momenta $l$ and $l\pm 2$ is in all partial waves quite reasonably described 
by the truncated OPE-potential. Thus, the channel coupling is easily 
controlled by the truncation parameter. For partial waves with  $l\ge 4$, the 
repulsive core is not required (for $E_{lab} < 400$ MeV) since the phase 
shifts with $l\ge $4 are completely determined by the long-range OPE tail of 
the interaction. Thus, the NN potential for even partial waves has the form:
\setcounter{equation}{32}
\begin{equation}
V_{NN} =V_{loc}+V_{sep},
\end{equation}
\noindent where $V_{loc}$ is a local $l$-{\em independent} part of
the interaction which, however, depends on spin and parity:
\begin{equation}
V_{loc}(r)=V_C(r)+V_T(r)\hat S_{12};
\end{equation}
\begin{equation}
V_C(r)=V_0\exp (-\eta r^2)+V^{OPE}_C(r)\cdot f_{tr}(r);
\end{equation}

\begin{equation}
V_T(r)=V^{OPE}_T(r)\cdot \left ( f_{tr}(r) \right )^n;
\end{equation}
where the cut-off factor is
\begin{equation}
f_{tr}(r)=1-\exp (-\alpha r).
\end{equation}
The exponent $n$ is different for various versions of the model.
The OPE potentials $V^{OPE}_C$ and $V^{OPE}_T$
have the standard form:
$$V^{OPE}_C(r)=V_0^{OPE}\exp (-x)/x;\qquad x=\mu r;$$
$$V^{OPE}_T(r)=V_0^{OPE}(1+3/x+3/x^2)\exp (-x)/x.$$
We use an averaged $\pi$-meson mass 
$\mu$ and a $\pi NN$ coupling constant $g^2_{\pi NN}/4\pi =13.8$ 
\cite{Kuk92,Kuk85,Kuk86a}
$$V_0^{OPE}=-{g^2_{\pi NN}\over 4 \pi}{\mu^3\over 4 M_N^2}=-10.69\, 
\mbox {MeV}, \; \mu=0.6995 \, \mbox {fm}^{-1}.$$
The separable repulsive core $V_{sep}$ has the form (for $l$=2 and 3):
\begin{equation}
V_{sep}=\lambda |\varphi ><\varphi |;
\end{equation}

\begin{equation}
\varphi=N r^{l+1} \exp \left ( -{1 \over 2} \left ( {r \over r_0} \right )^2 
\right );  \; \; \; \int\varphi^2 dr=1.
\end{equation}

The strength constants $\lambda$ for the separable core vary for different
$J$ and $l$, whereas the radius of the repulsive core $r_0$ varies 
only slightly from state to state (see Table 3). For $l\ge 4$ the 
term $V_{sep} $ is absent. Parameters for several variants of the 
$NN$-potential differing in the truncation of the tensor part in 
triplet-even channels are given in Table 3.
We have found (see Sect.4) that the properties of NN- and 3N-systems calculated
with M-type potentials (this is also true for conventional NN potentials
\cite{Mac89,Wit91,Str88,Glo95,Ish87} strongly depend on the behavior 
of the tensor potential at small distances. In the 1986 version of Moscow 
potential (variant (A) in Table~3), an exponential truncation 
of the tensor part (see eq.(37)) with a value $n$=3 and an inverse radius of 
truncation $\alpha =3\mbox{ fm}^{-1}$ was used. In that case, the tensor 
potential truncated according to eqs. (36) and (37) corresponds to a large 
negative constant at the origin ($V_T(0) =-5500 $ MeV).  Therefore, there is 
a very strong coupling of the $^3S_1-{ }^3D_1$ channels at small distances. 
This coupling results in a large loop in the deuteron
$D$-wave, as well as in a large $D$-state admixture $P_D$  of 6.4\%. The 
relative amplitude of this loop $A^D_{loop}$ compared to the maximum value of 
the $D$-wave function is given in Table~3. 

However, in view of the relatively weak quark-quark tensor force, it is
desirable to have an interaction variant, for which the node in the deuteron
$D$-wave is absent. Such a variant was found in our previous work \cite{Kuk92},
where a step factor:
\[f_{tr} =\frac {(pr)^6}{1+(pr)^6}\]
has been used for the truncation. The step-like truncation 
leads however to technical difficulties when used in a three-body calculation. 
Therefore, we present here some alternative variants of the triplet-even 
potential which still have an 
exponential truncation factor $(1-\exp (-\alpha r))^n$ with $n>3$, for which
the relative magnitude of the $D$-wave loop varies from 0.5 down to 0.02. 
It is important to note that the stronger the truncation of the 
tensor potential, i.e. the weaker the tensor interaction 
at small distances, the larger the 3N binding energy (see the subsequent 
section).

The dependence of all observables on the truncation factor of the central part
$V^{OPE}_C$ is very weak. For example, a stronger truncation in $V^{OPE}_C $ 
is easily compensated by a corresponding strengthening of the central 
attractive (Gaussian) part of the interaction. Therefore, in the present 
three-body  calculations, we have used only one of the new variants for the 
singlet potential presented in Table 3 (namely, variant c).
In addition, we have found that a fit to the experimental 
phase shifts in the interval 0-400 MeV does not uniquely determine 
the width $\eta $ and depth $V_0$ of the deep potential in the central 
part of the NN interaction (35). Any variation of the width $\eta $ can 
always be compensated by an appropriate variation of the depth $V_0$, without 
changing the quality of the phase shift fit. As a result, it is possible to 
choose the same width for the triplet $^3S_1-{ }^3D_1$ ($\eta_t$)
and the singlet $^1S_0-{ }^1D_0$ ($\eta_s$) channels. Alternatively, it is 
possible to choose the same values of the depth $V_0$ in triplet and singlet 
channels. In this case the width $\eta_s$ and $\eta_t$  would be different.

We emphasize that the equality $\eta _s=\eta _t$ is impossible to reach in
simple conventional potential models \cite{Bro76}. In other words, 
the width of the triplet potential is always different from that of the 
singlet potential. Both parameters (i.e. width and depth) of the 
conventional potential model are {\em uniquely} determined by the values of 
the low-energy effective range parameters, i.e. by the scattering length and 
effective range. It was recently found \cite{Dub96} that owing to some 
freedom in the choice of the three main parameters of the interaction, 
i.e. width 
$\eta$, depth $V_0$, and the truncation radius of the OPE potential 
$\alpha $, the values of the width parameters $\eta $ and $\alpha ^2$ can 
be taken to be equal, i.e. $\eta =\alpha ^2$. We then obtain the simplest
two-parameter model of the Moscow potential, where only two parameters
(width and depth) are fitted to the scattering length and effective range
of the $^1S_0$ partial wave. However, unlike conventional force models 
(see ref. \cite{Bro76} Chap.2), 
we obtain with this simplest two-parameter model 
an excellent description of the 
$^3S_1-{ }^3D_1$ and $^1S_0$-wave phase shifts in a large interval 
of 0--500 MeV instead of a good fit only in the range 
between 0--15 MeV typical for 
conventional models. Simultaneously, we obtain a  good description of the 
main static properties of the deuteron \cite{Dub96}, including the quadrupole moment, 
charge r.m.s. radius and the asymptotic normalizations $A_S$ and $A_D$.

We attribute this success to the choice of the proper degrees of freedom for a 
potential description. Conventional models try to describe both,
the short-range six-quark and the 
long-distance one-boson exchange aspects of the NN interaction. This leads to 
a complicated energy and momentum dependence of the resulting NN potential.
We claim that the Moscow potential model is so simple, 
because we do not try to cast the six-quark aspects of the NN 
interaction into the same formalism as the asymptotic long-range part.
The six-quark aspects in the NN sector are more adequately described by an 
orthogonality condition, rather than by a local repulsive NN potential.

The parameters of the Moscow potential model are listed in Table~3.
The deuteron wave function calculated with model B is shown in Fig. 4,
and its static properties are listed in Table 4. 
In Fig. 5, the phase shifts of the 
{\it even} partial waves for variant (B) are compared with the 
experimental ones \cite{Arn92}. 
The description of the NN phase shifts is for all variants 
listed in Table~3 quite reasonable.
However, our description of the  
mixing parameter $\epsilon_1$
especially for energies $E > 400$ MeV, is not yet satisfactory.
The mixing parameter $\epsilon_1$ is determined by the 
value of the matrix element
$\langle \Psi_S\vert V_T \vert \Psi_D \rangle $.  
Our overestimation of $\epsilon_1$ at higher energies 
is a consequence of the behavior of the $^3S_1$ and $^3D_1$ 
eigen functions at short distances. While the $S$-wave
has a node and a rather pronounced loop, there is almost
no loop in the $D$-wave. This leads to a rather sharp increase  
of $\epsilon_1$ at higher energies.
\footnote{In contrast to this,
in the first version of the Moscow potential model \cite{Kuk85},
there were two coinciding nodes and loops in both the $S$ and $D$ waves 
and a rather satisfactory fit to $\epsilon_1$ was obtained.}
In the more complete two-channel model, the short-range
$S$-wave loop should be significantly reduced and we expect
a flatter behavior of $\epsilon_1$ in better agreement
with modern data.

\subsection{Exclusion of the deep-lying bound state}
\label{exclusion}

\par
The "extra" bound states in $S$- and $P$-waves have to be 
excluded if the potential is used in few-nucleon calculations. 
As was already discussed in Section 2, this is most conveniently done 
by means of the orthogonalizing
pseudo-potential (OPP):
\begin{equation}
V_{OPP} =\lambda \Gamma, \qquad \Gamma =| \varphi _f> <\varphi _f|
\end{equation}
with a large positive value of the strength constant 
$\lambda $ \cite{Kuk78,Kuk83a}. 
In practice a value of $\lambda \sim 10^5-10^6$ MeV is quite
sufficient. In the case of coupled channels $^3 S_1-{ }^3D_1$ and $^3P_2-{ }^3
F_2$ the function $\varphi _f$ in (40) has two components. Therefore,
it is necessary to use  a two-channel {\em
matrix } projector of the form:
\begin{equation}
\hat {\Gamma }_2=\left (
\begin{array}{cc} | \varphi ^1_f><\varphi
^1_f| & | \varphi ^1_f><\varphi ^2_f| \\
| \varphi ^2_f><\varphi ^1_f| & | \varphi
^2_f ><\varphi ^2_f| \\
\end{array} \right )
\end{equation}
where $|\varphi ^1_f>$ and $|\varphi ^2_f>$ are the upper and lower 
entries of a two-component column vector, corresponding to the
"extra" bound state.
However, the matrix projector considerably complicates three-body
calculations. Therefore, we use instead a one-component projector:
\begin{equation}
\hat { \Gamma } _2\to \hat { \Gamma } _1=\left (\begin{array}{cc} |
\tilde {\varphi }_f><\tilde {\varphi }_f| & 0
\\
0 & 0 \\
\end{array} \right ),
\end{equation}
acting on the  "main" channel, i.e. on the channel possessing highest weight
in the "extra" state. In other words, in $^3S_1-{ }^3D_1$ channels 
we use only an $S$-wave projector.
%and in the $^3P_2-{ }^3F_2$ channels only a $P$-wave projector. 
In this case the function $\tilde 
{\varphi}_f$ is selected so that the inclusion of the above 
one-component projector into the Hamiltonian gives a result that 
is as close as possible to the action of the {\em two-component} 
eigenprojector $\Gamma _2$.
Strictly speaking, the replacement of the matrix projector (41) by a 
single-channel
projector (42) should cause some modification of the appropriate NN phase 
shifts. We have found, however, that this replacement influences the 
description of the phase shifts in coupled channels only very weakly. 
Moreover, employment of the simple Gaussian function with suitable 
values of $r_0$ in $\tilde {\varphi}_f$ can ensure a quite reasonable and 
accurate description of scattering phase shifts.

We will show, in particular, how the approximate $S$-wave projector $\Gamma
=| \tilde {\varphi }_f><\tilde {\varphi }_f| $ acts in the coupled
$^3S_1-{ }^3D_1$ channels.
The dependence of the first two energy levels on the
orthogonalizing coupling constant $\lambda $ for the 
$^3 S_1-^3 D_1$ channels is shown in Fig.6(a-b) for the potential
(B). If the projector is absent, there are two bound states: the 
"extra" ground state with energy $E_0=-599$ MeV with a $D$-wave
probability of 6.3$\%$, and the second bound state describing the 
physical deuteron with $E_1=-2.2245$ MeV and a D-wave probability 
of $P_D$=5.75 $\%$. If the exact two-component state
$\varphi _0$ is used in $\Gamma_1 $, the energy of the ground state $E_0$ is
shifted by an amount $\lambda $ when $\lambda $ is varied, whereas the
deuteron energy $E_1$ does not vary at all (by definition). The deviation of 
the curves $E_0 (\lambda )$ and $E_1 (\lambda )$ from the straight line 
characterizes the difference between the application of the 
approximate (single-channel)
$S$-wave projector and the exact matrix eigenprojector $\hat {\Gamma }_2$.
As one can see, this difference is only important in the narrow range of
values $\lambda =685-690$ MeV, where the energies $E_0$ and $E_1$ of both
states come close to each other and the states are mixed strongly due to
approximate orthogonality of $\tilde {\varphi }_f$ and $\varphi _1$ (their
overlap being $<\tilde {\varphi }_f|\varphi _1>=0.001$). 
For $\lambda > 690$ MeV,
the deuteron state becomes the new ground state and its energy grows very
slowly with $\lambda $ approaching the limiting value $E(\infty )=-2.2246$
MeV. On the other hand, the original ground state energy $E_0(\lambda)$ 
continues to rise and finally transforms into a narrow resonance. In the 
case of the exact eigen projector, it would be transformed into a bound 
state embedded into the continuum with the positive energy $E_0+\lambda $. 
This resonance strongly distorts the phase shifts only in a narrow energy 
range near $E_{cm}=E_0 + \lambda $. 
But since the values for $\lambda$ used in the present three-nucleon
calculation are of the order of 
$\lambda \sim 10^6$ MeV, this distortion is far away from the 
physically relevant region.
It is worth to note that if one takes only the $S$-component of the 
{\em exact} wave function $\varphi _0$ as $\tilde {\varphi }_f$ in the 
single-channel
projector (42), the result turns out to be much worse. In other words, the
result of the application of the projector $| \varphi ^{(1)}_0><\varphi
^{(1)}_0| $, i.e. {\em using the exact} first component of
two-component vector, differs much more from the matrix 
eigenprojector
$\hat{\Gamma }_2$, than the {\em approximate } projector $|
\tilde { \varphi }_f><\tilde {\varphi }_f| $. This is explained by the fact
that the orthogonality condition which is automatically satisfied for the
eigenprojector ${\hat \Gamma}_2$ is:
\begin{equation}
< \Phi _0| \Phi _1>=
<\varphi ^{(1)}_0| \varphi ^{(1)}_1>+<\varphi ^{(2)}_0| \varphi ^{(2)}_1>=0.
\end{equation}
From here it does not follow, in general, that $<\varphi ^{(1)}_0| \varphi
^{(1)}_1>=0$. At the same time it is possible to make the overlap of the
deuteron with some approximate function $\tilde {\varphi }_f$ 
arbitrarily small. Then the energy $E_1$ will not depend on $\lambda $, 
as in the case of the application of the exact eigenprojector
${\hat \Gamma}_2$.
However, it is impossible to choose a simple approximate function that will be 
orthogonal to the scattering wave functions of the continuous spectrum
for {\em all} energies. Therefore, due to the employment of the 
approximate projector, the scattering phase shifts are inevitably distorted 
in some small energy interval.

Bearing in mind the use of the potential in the three-nucleon system, 
where large values of the constant $\lambda $ are needed, we restrict 
ourselves to the
approximate projector with a simple Gaussian form factor.
Then the additional pseudo-potential $V_{sep}$ has the form of eq.(40) in all 
channels. For $S$- and $P$ partial waves, the constants $\lambda$ 
have to be sufficiently large ($\sim 10^5-10^6$ MeV) in order to ensure the 
exclusion of "extra" bound states, while their values in $D$-waves are 
determined by fitting the experimental phase shifts. The corresponding 
values of $r_0$ for the approximate projector are also given in Table~3. 
In the $^3 D_1$ channel and in all higher partial waves with $l>3$, the term 
$V_{sep}$ is absent.

\section{Properties of the three-nucleon bound state}
\label{properties}

\par
Here, we discuss the predictions of the new versions of
the Moscow model described in section 4. The 3N calculations
have been done
by a variational method on a very large, nonminimal and nonorthogonal Gaussian
basis \cite{Kuk77,Kra77} and are, in general, rather similar to the 
very accurate
3N-calculations of Kameyama et al. \cite{Kam89}. With this basis
the {\em absolute} accuracy for the eigenenergy 
is about 100 keV. The {\em relative} accuracy for the Coulomb
displacement energy is even higher.

As follows from many previous 3N bound state calculations, 
the total contribution of
odd partial wave interactions to the 3N binding energy is small and does not 
exceed 0.2 MeV \cite{Had85}. We emphasize 
that odd partial waves have 
been included in our variational basis for the 3N bound state
calculation but we did not take into account 
the contribution of the odd partial waves in the NN force. 
In the NN force,
we include all {\em even} partial waves up to total orbital angular momentum 
$L=4$.
Our results are presented in Table~5, where we list several static properties
of $^3$H and $^3$He such as binding energies $E$, root-mean-square charge
radius $r_{ch}$, the percentages of $D$- ($P_D$) and $P$-waves ($P_P$), and the
Coulomb displacement energy $\Delta _{Coul}$ for the difference
$E_B({ }^3H)-E_B({ }^3He)$. In addition, we list 
the average kinetic energy $T$ in the 
ground state, as well as certain characteristics of the $^3$H charge form 
factor, i.e. the position of the first minimum $q_{min}$, the second maximum 
$q_{max}$ and the value of $|F_{ch}|$ in the second maximum $F_{max}$.

\subsection{Dependence on the projection constants}
\label{dependence}

\par
First, we discuss the dependence of the three-nucleon 
properties found with the above force model on the projection constants 
$\lambda $.
In Fig. 6(a-b) we show the dependence of the 3N ground state energy $E_{3N}$ 
on the value $\lambda (^3S_1)$ multiplying the projector $\Gamma$, projecting 
onto an "extra" bound state in the $^3 S_1$ channel. The 
two-body state energies 
$E_0 (\lambda)$ and $E_1(\lambda) $ shown in Fig.~6 were already 
discussed in Section 4. All other
parameters of the potentials are fixed, including the constants 
$\lambda $ for the 
other channels. For $\lambda < $680 MeV the 3N-system remains
unbound and the variational value $E_{3N} $ practically coincides
with the energy of the two-body ground state $E_0$. When 
$\lambda$ is increased beyond 682 MeV, the system becomes bound with 
$E_{3N}=-6$ MeV. If $\lambda$ is increased further the three-nucleon 
binding energy decreases very slowly approaching its limiting 
value $E_{3N}(\infty) =-5.74$ MeV, because these test calculations were
done in a comparatively small basis. The saturation occurs for $\lambda
=10^5-10^6$ MeV. For higher values of $\lambda $ the numerical results 
become unstable. The convergence of variational calculations  with 
respect to $\lambda $ was 
investigated in some detail in Ref. \cite{Vas88}. 
The physical deuteron becomes
the ground state of the two-nucleon subsystem for $\lambda >$688 MeV when
$|E_{3N}|$ is already less than $6$ MeV.

These results clearly show that it is impossible
to fit the binding energy of the three-body system to the experimental value
by adjusting the parameter $\lambda $. It is interesting to note that 
the system becomes bound when the repulsion in pair subsystems
increases\footnote{Certainly, the absolute value of three-body binding
energy decreases in the process, however the energy of the two-body threshold
decreases still more quickly.}. The reason of this interesting phenomenon is
the following. When the projection constant in the singlet channel is 
chosen as e.g. $\lambda_s =10^6$ MeV and the one in the triplet 
channels as e.g. $\lambda_t << 10^6$ MeV (and varying), 
the almost complete space symmetry of the 
three-nucleon system is broken, and the weight of wave function components 
with mixed orbital permutational symmetry $[21]_X$ is increased.

On the other hand, if the constants $\lambda (^3S_1) $ and 
$\lambda (^1S_0)$ evolve simultaneously, we find a different behavior 
${\tilde E}_{3N}(\lambda )$ that is shown by the dashed line in Fig. 6b. 
Now, the 3$N$ system remains 
bound for all $\lambda $. However, for the allowed values of $\lambda $ 
that is when both "extra" bound states in the triplet and singlet channels 
are pushed sufficiently high above the physical state, the three-nucleon 
binding energy $|E_{3N}|$ appears at a value  
less than $7$ MeV. So, in this case it is also impossible to fit the 
three-nucleon binding energy by shifting the "extra" two-body state to 
the positive energy region. Therefore, all results for the three-nucleon
system discussed below have been calculated
with sufficiently large values $\lambda_s$ and 
$\lambda_t (\sim 10^5 - 10^6
$ MeV) for which complete saturation has been reached, and the results are, 
in some sense, $\lambda $-independent.

\subsection {Binding energy of $^3$H and $^3$He.}
\label{binding}

\par
All variants of the model studied here differ only in the form
of the truncation factor for the OPEP-tensor force at small distances 
and lead to overall similar results (see tables 3,4,5) except for one 
previous version of our model \cite{Kuk86a} (variant A in Tables 3-5).
In the course of the calculations we have found that the stronger
cut-off in OPEP tensor force, the larger the 3N
binding energy (see the Tables 3 and 5). In this respect our results 
are very similar to those found by Sasakawa
and Ishikawa \cite{Ish87}. In their 3N studies with the Reid Soft-Core (RSC) 
NN potential using various truncation factors of the tensor force 
they found that when the tensor force is cut off
more strongly at short distances (and the central attractive part is 
correspondingly increased in order to keep the deuteron binding energy 
invariant), the 3N binding {\em rises} as well.

However, there is an important difference between the present results 
and those
in Ref. \cite{Ish87}. In our case, the partial replacement of the
short-range tensor force by a central force 
does not lead to a noticeable reduction of the tensor mixing parameter
$\varepsilon _1$. In fact as the truncation is increased, 
the mixing parameter does not tend to zero, but, instead of this, it rises 
more sharply for larger $n$.  
This is related to the essentially different
character of interference between tensor and central force (both in 2N and
3N systems) in our case as compared to the traditional nuclear force models
(see the last paragraph in Sec.\ref{onechannel}). 
However, we do not expect drastic effects of our $\epsilon_1$ 
description on the 3N binding energy, at least not within
the corridor given by the different versions of our model.
In fact, the difference in $\epsilon_1$ values between the RSC and 
the Paris potential model is of the same size as that between the present 
version and experiment \cite{Bro94}.   

The $^3$H binding energies obtained in the present study ($\sim $6.05 MeV)
for the two most realistic variants of the potential (B and C) are much higher
than the value of $\sim $4.5 MeV obtained by Hahn et al. \cite{Hah86} in
their first 3N calculation with the earlier version of the Moscow potential.
This large difference is caused by the following reasons:

(i) In the Hahn et al. \cite{Hah86} calculation and also in the early 
1983 version of the Moscow potential  an old version of the singlet $^1S_0$
potential fitted to the 1983 phase shift analysis \cite{Arn83} was employed.
Our new singlet potential is fitted to the SAID database of 1995 \cite{Arn92} 
and resulted in some increased 3N binding (about $\sim $0.2 MeV).

(ii) Another, more important reason for the present improvement
of the 3N results is the use of a stronger
truncation of the short-range OPEP tensor force and the respective
reduction of $P_D$ in the deuteron and 3N systems - see the versions B, C
and D in Table 4. The sharper truncation removes or supresses to a 
significant degree the inner node and the respective loop in the 
$D$-wave at short distances.

(iii) A very restricted configuration space of only 5 basic channels 
$^3S_1-{ }^3D_1$ and $^1S_0$ was employed in the Faddeev calculation 
of Hahn et al. \cite{Hah86} and the contribution of all $l>$2 higher 
partial waves 
was neglected. The latter contribution are quite important in our case 
especially in the Faddeev like 3N-approach.

We have previously emphasized the extremely important contribution of 
higher partial waves as manifested by a sharp increase of the average 3N 
kinetic energy. Compare the third column in Table 5 with the corresponding
results \cite{Mac93,Wir91}
 for traditional force models, in particular for model B: 

\[ E^{3N}_{kin}(\mbox{Paris})=42.6\; \mbox {MeV};
\quad E^{3N}_{kin}(\mbox{RSC})=49.3\; \mbox {MeV};\]

\[ {\rm while} \ \ \  E^{3N}_{kin}(\mbox{Moscow})\simeq 150\; 
\mbox {MeV}, \]
is more than in three times higher.
It is worth to recall here that the modest 20$\%$ increase in the average 
kinetic energy in the RSC potential as compared with the Paris potential, 
results already in an enhancement of the contribution of higher partial waves 
to the binding energy by as much as a factor of 2.5. Thus, one can easily 
imagine how large a contribution from higher partial waves may result in our 
case. A more detailed analysis of these results will be carried out elsewhere. 

We must keep in mind that, contrary to 
conventional force models, our NN-potential corresponds only to a  
one-channel approximation of the complete two-phase model, that is
by definition of the NN potential, only the effective NN channel with 
unexcited nucleons. As was argued above, the effective channel potential is 
phase-shift equivalent to the initial two-phase interaction model of hybrid 
type (see eqs. (19) and (24)) and the replacement of the local repulsive core 
by a two-channel interaction model (e.g. as in case of QCB model 
by Simonov \cite{Sim84}) will lead to some additional binding of the order of 
some $\sim 0.8\div 0.9$ MeV \cite {Bak90}, simply because of 
the {\em presence} of the second channel.
There is an additional reason for extra-binding in 3N- and generally
in many-nucleon systems if the full two-phase interaction model is 
considered.  One has to take into account the 
interaction between the six-quark bag and the spectator nucleon. 
This dynamical 6q-N channel coupling can be considered as some kind of 
three-body force that will also increase the 3N-binding energy. We expect, 
that these effects will mainly
change the {\em absolute} values of the energies while the {\em relative} 
energies, which will be discussed in the following, will be 
influenced only weakly. This is because the weights of the 
6$q$-bag components in $^3$H and $^3$He are small and have almost 
equal magnititude.

\subsection{Coulomb displacement energy, charge radii and form factors.}
\label{Coulomb}

\par
As follows from the results displayed in Table 5, the Coulomb
displacement energy:
\[\Delta _{Coul}=E_B(^3H)-E_B(^3He)\]
(for variants B and C with $n$=5 and 7 for the OPEP tensor force truncation) 
equals:
$$ \Delta _{Coul}\simeq 670 \mbox { keV} $$
for $E_B(^3H)$=6.05 MeV. For a more careful test of the model, it is important,
however, to find the value of $\Delta _{Coul}$ when $E_B(^3H)$ is scaled to
its experimental value $-8.48$ MeV since the larger the 3N binding energy
the smaller the r.m.s. charge radius and the higher the Coulomb energy
of $^3$He.
Thus, we slightly increase (by $\sim 2\%$) the strength of the central
(triplet and singlet) potential well to fit the experimental value of
$E_B(^3H)$, and for this case we analyze the Coulomb energy of $^3$He and
the r.m.s. charge radii of $^3$H and $^3$He with our model (see the two last
rows in Table 5).

The extrapolated Coulomb displacement energy turns out to be:
$$ \Delta ^{extra}_{Coul} =737 \mbox{ keV} $$
which is only by 27 keV less than the experimental value. Here, we note
in passing that in the full two-phase model calculation the magnitude of
$\Delta _{Coul}$ would increase further because the second channel 
comprises the more tightly packed 6$q+N$ configuration. 

The most likely reason for this improvement in the prediction of 
$\Delta ^{extra}_{Coul}$ is the appearance of the 
inner radial nodes and corresponding short-range loops along every
interparticle $r_{ij}$ coordinate (and also along both Jacobi coordinates)
in our 3N wave functions. 
The inner loops along the $r_{ij}$ coordinates have only a minor influence on
the r.m.s. radii of the charge or matter distribution 
because the latter are determined mainly by {\em one-particle} coordinates 
$r_i$ (as measured from the center of mass).
Contrary to this, the Coulomb potential between two protons in $^3$He is
maximal just at zero interparticle separation $r_{ij}$ where we have
some enhancement of particle density due to the presence of the inner loops 
in our wave functions.

This naturally explains why conventional force
models leading to a short-range suppression of the wave functions along
every interparticle distance miss some 100 keV in the Coulomb
energy of $^3$He, whereas our model, due to the inner loops leads 
to a prediction that is close to the experimental value. Although
our result is still qualitative in the sense that it involves an extrapolation
to the empirical binding energy, the observation that the inner wave function 
loops are extremely important for the Coulomb displacement energy is likely 
to survive. Moreover, if the force model presented here is  
corroborated in subsequent studies, it would suggest that inner loops 
are ubiquitous in nuclear wave functions.

The recent quantitative explanation of the Coulomb displacement 
energy in the literature \cite{Pud95} using a force model with
a repulsive core is based on the assumption of a rather strong
charge symmetry breaking (CSB) term $V_{CSB}$ (difference between 
np and nn strong interaction). With $V_{CSB}$ added to the
the Argonne $V_{18}$ potential, one can almost completely explain
the 100 keV difference in $\Delta_{coul}$ between $^3He$ and $^3H$.
However, the same $V_{CSB}$ leads to a strong overestimation of
the Coulomb energy difference in the $^6Be-^6He$ system.
Therefore, it is conceivable that the contribution of the charge
symmetry breaking term $V_{CSB}$ in Ref. \cite{Pud95} 
is actually smaller. In any case, it is evident from the
discussion above that the assumption of a strong charge symmmetry 
breaking interaction $V_{CSB}$ is not the only mechanism  
that can explain the 100 keV Coulomb energy difference in the
3N system.        

The values of the extrapolated charge radii of $^3$H and $^3$He 
(see last row in Table~5) turn out to be larger than the 
corresponding experimental values.
However, it should be emphasized once again that the 3N-calculations
with one-component force model presented here, i.e. carried out only in
the NN (or 3N-) sector, corresponds just to the projection of the total
two-phase model wave function onto the pure 3N-component. 
The two-phase solution includes also compact 6$q+N$ components
with a sizeable probability,
which may considerably reduce the 3$N$ charge radius. Similarly, the full 
two-phase model will lead to important modifications in the charge form 
factors of $^3$H and $^3$He. Our results for the $^3$H charge form factors 
calculated
with the one-phase Moscow potential are displayed in Fig. 7 together with 
the predictions for the extrapolated (to experimental $E_B(^3H)$) solution. 
We see that the behavior of the form factor for $q^2>14 \,\mbox{fm}^{-2}$ 
is rather similar to predictions calculated with a conventional force 
model such as the RSC potential. However, we expect appreciable
modifications for $q^2>14 \, \mbox{fm}^{-2}$ for the original two-phase model, 
in which the sign of the 6$q+N$ component is {\em opposite} to that of the 
main 3$N$-component \cite{Obu91}. According to refs. \cite{Bak90,Nam82} 
this fact should 
considerably improve the description of the 3$N$ charge form factors in the 
secondary maximum.

Summarizing we conclude that the description of both the 3$N$ binding energy 
and some important 3$N$ observables will presumably improve when passing from
a one-phase potential model to a two-phase model due to the 
explicit appearance of the 6$q$+N component in the three-body solution.

\section{Conclusion}
\label{conclusion}

\par
The force model presented in this paper -referred to as 
M-type model- differs in a few important aspects from traditional 
NN interaction models currently in use.

First, M-type models necessarily incorporate additional
orthogonality condition(s) with respect to certain nodeless functions
$\varphi_0$.
The orthogonality condition can be interpreted as the projection of 
six-quark compound states, $\varphi_0$ with maximal space 
(or permutational) symmetry  onto the asymptotic NN-channel.

Second, M-type models are characterized by the presence of a
deep attractive well while a part of the short-range repulsive core (or the
whole core) is replaced by appropriate orthogonality
conditions\footnote {The M-type models presented here are differ from
the OCM approach by Saito \cite{Sai69} in a few essential aspects: our model 
is 
based on another, more convenient mathematical formalism that is used to 
incorporate the additional orthogonality conditions. Second, the physical 
meaning of the "forbidden" subspace is quite different in our case because 
we start from a two-phase interaction model. Third, we generally combine 
the (reduced) repulsive core and the additional orthogonality
condition.}. The consequences of the different 
off-shell behavior of conventional and M-type NN force models
can be tested in NN bremsstrahlung \cite{Neu97}.  

Finally, the models presented
here have a physical interpretation that is rather different from that
of conventional force
models. A standard force model aims at describing the NN interaction in
the total configuration space in the form of some potential.  
Contrary to this, we start from the assumption that certain six-quark compound
states of maximal permutational symmetry are present in the lowest partial
waves. The latter have only very small projections onto the 
non-antisymmetrized asymptotic NN channel. However, at the same time they 
are strongly coupled by the six-quark Hamiltonian with the clusterized 
NN channel at small distances $r<$1 fm. 
As a result, these compound states cannot be
described by {\em any} reasonable NN potential. 
These states can only be described by a very complicated nonlocal and 
energy-dependent NN interaction operator.
However, even when utilizing such
complicated interaction operators \cite{Ord96} we are forced to introduce many
adjustable or free parameters to fit the data. Thus, due to the formation
of the above compound six-quark states, the whole system cannot be
described by any simple NN potential. In order to get an efficient
potential description, it is necessary to remove the above 
{\em nonpotential pieces} from the full interaction operator. 
This can be done with the help of the two-phase orthogonalized
model described in Sect.2. The exclusion of six-quark compound states 
can be conveniently accomplished by the well known Feshbach formalism. This 
eventually leads to the Moscow type models with its two orthogonal channels.

Our approach makes it possible to describe rather accurately
both NN phase shifts up to 500 MeV, and the deuteron structure with a 
truncated one-pion exchange potential together with a 
simple deep {\em static} potential. Altogether we use 6 parameters
with clear physical meaning.
This evidently confirms 
the usefulness of the present two-phase approach. Other hybrid models 
which combine both quark- and meson exchange degrees of 
freedom make use of {\em nonorthogonal} quark- and meson exchange channels, 
i.e. some mixture of both. As a result, they contain {\em nontrivial} 
energy dependence and nonlocalities. In addition,  most hybrid models 
do not offer a microscopic interpretation of the NN channel.
 
A clear separation of the nonpotential pieces of the NN interaction
and the subsequent parametrization of the rest in an orthogonal 
{\em subspace} is the main physical idea underlying our approach. Therefore, 
the success of the model may be taken as some evidence for the 
formation of six-quark compound states (i.e. dibaryons) at short-range in 
low partial waves. As a result of the elimination of 
6q-compound states, the 3N-, 4N- and other many-nucleon systems are 
underbound. 

In this work we have used only the one-phase {\em effective} two-body 
potential component of the complete two-phase model.  
In the future, 
it would be interesting to take the next step and to 
incorporate the second, i.e. 6$q$+N component in few- and many-body
calculations. In addition, it would be interesting to include vector 
mesons into the theoretical description. Due to the additional 
orthogonality condition, their effect will 
not be as large as in standard force models, where the 
$\omega$NN coupling constant  is in contradiction with the
SU(3) prediction \cite{Mac89}. In our case we are able to keep the  
$\omega$NN coupling constant as low as SU(3) predicts.

\bigskip
\bigskip
\bigskip

\centerline{\large \bf Acknowledgments.}

We are thankful to many of our colleagues in Moscow and T\"ubingen for 
long term fruitful discussions on the topics of the present study, 
especially to Profs. E. W. Schmid, V.G.Neudatchin, and Dr. I. T. Obukhovsky. 
One of the authors (V.I.K.) is very grateful to Profs. Steven Moszkowski and
Gerry Brown for many useful comments on our approach. He is also grateful 
to Prof. C.N.Yang, Head of Theoretical Physics Institute at Stony Brook, for 
his invitation to the Institute.
The Russian authors thank the Russian Foundation for Fundamental
Research (grant No.96-02-18071) and the Deutsche Forschungsgemeinschaft 
(grant No. Fa-67/20-1) for partial financial support.

\newpage

\centerline{\large \bf Figure captions}

\bigskip

{\bf Fig. 1.} The scattering wave functions at $E_{lab}$=1 MeV for three
$S$-wave phase shift equivalent potentials: initial local MT-V (curve 1), the
intermediate nonlocal potential with $r_0$=0.2 fm (curve 2), and the purely
attractive potential with the projector close to the eigen projector with 
$r_0$=0.4 fm (curve 3).

\bigskip

{\bf Fig. 2.} NN $S$-wave phase shifts for center of mass energies 
between 0-400 MeV for the
initial MT-V potential (dashed line) and for two modified potentials $V'(r)$
fitted to $S$- and $D$-wave phase shifts (see Table 2): (i)  with $r_0$=0.15
fm (solid line) and (ii) with $r_0$=0.2475 fm (dotted line).

\bigskip

{\bf Fig. 3.} $D$-wave phase shifts for the same three variants of the
interaction potential as in Fig.2.

\bigskip

{\bf Fig. 4.} The deuteron S-wave (l=0) and D-wave (l=2) functions for 
model B. The deuteron
wave functions calculated with the Paris potential \cite{Lac81} 
are shown for comparison.

\bigskip

{\bf Fig. 5.} The even parity phase shifts for the new version of the $NN$
Moscow potential (variant B) are compared with the data of the 
energy-dependent phase-shift analysis 
by Arndt et al. \cite{Arn92} (triangles, dotted line for $\varepsilon _1$). 
For $\varepsilon _1$  we also show the Nijmegen phase-shift analysis 
PWA93 \cite{Nij93} (dashed line) and the single energy phase-shift analysis 
by Arndt et al.\cite{Arn92} (circles).

\bigskip

{\bf Fig. 6.} (a) Effect of the approximate $S$-wave projector on the 
discrete spectra of 2N and 3N systems. $E_0$ and $E_1$ are the 
energies of the first two 2N-states, $E_{3N}$ is the energy of 3$N$ 
ground state, $\lambda $ is the $^3S_1$-wave projector coupling constant. 
(b) The dotted line shows the dependence of the 3N-bound state 
on $\lambda $ for the case $\lambda _s=\lambda _t$.
\bigskip

{\bf Fig. 7.} The charge form factor of $^3$H for different versions of the
Moscow NN potential: the previous version (A) (short-dashed line), the
present version (variant B) (dashed line), and the potential
extrapolated to $E_{exp}({ }^3$H) (solid line). Also are shown are 
the previous ($\Box$) and the new \cite{Amr94} ($\triangle$) 
experimental data. 

\newpage

\oddsidemargin=0 cm
\begin{table}[h]

\caption[MTV]{The parameters of the modified potentials reproducing the
$S$-wave phase shifts for the initial MT-V potential with different values of
the "projector radius" $r_0$ and properties of the wave function.}
\bigskip
\begin{tabular}{|c|r|r|c|c|c|r|c|} \hline
$r_0$, [fm] & $g'_R$ [MeV] & $g'_A$ [MeV] & $r_{node}$ [fm] & $\beta $ &
$\chi ^2 (deg^2)$ & $\varepsilon _0$ [MeV] & $\tilde {\varepsilon }_0 
$[MeV] \\ 
& & & &(1 MeV) & per point & & \\
\hline
0.00 & 1458.1 & -578.1 & 0.00 & ---& --- & -0.4136 & --- \\
0.15 & 992.0 & -511.0 & 0.25 & 0.03 & 0.40 & -2.51 & -0.385 \\
0.18 & 673.8 & -468.6 & 0.28 & 0.06 & 1.07 & -7.55 & -0.370 \\
0.20 & 433.0 & -438.0 & 0.30 & 0.07 & 1.70 & -36.06 & -0.364 \\
0.25 & -5.0 & -389.0 & 0.35 & 0.16 & 2.50 & -430.30 & -0.399 \\
0.30 & -63.6 & -403.0 & 0.40 & 0.28 & 1.60 & -683.70 & -0.611 \\
0.40 & -106.2 & -494.0 & 0.50 & 0.47 & 0.05 & -528.00 & -0.919 \\
0.50 & -287.2 & -582.5 & 0.60 & 0.63 & 0.10 & -378.80 & -0.880 \\
\hline
\end{tabular}

\bigskip

\caption[Variants]{Variants of the MT-V potential with different 
values of the "projector radius" $r_0$, fitted to $S$- and $D$-wave phase 
shifts of the original MT-V potential and the two-body bound state energy 
$\varepsilon _0$=-0.4136 MeV, together with the corresponding results for the 
three-body calculations.}
\bigskip
\begin{tabular}{|r|r|r|r|r|c|c|c|c|c|c|}
\hline
$r_0$ & $g'_R$ & $g'_A$ & $E_3$ & $T_3$ & $\%$S &
$\%$D & $^{\approx }E^{rep}_3$ & $q_{min}$  & $q_{max}$ &
$F_{max}$ \\
fm & MeV & MeV & MeV & MeV & & & MeV & fm$^{-2}$ & fm$^{-2}$ & \\
\hline
MT-V & 1458.05 & -578.09 & -8.25 & 30.6 & 99.0 & 0.75 & -7.52 & 19 & 24.0 &
4.8$\cdot 10^{-4}$ \\
0.15 & 999.36 & -514.00 & -8.42 & 40.3 & 98.5 & 1.1 & -6.73 & 19 & 24.0 &
5.4$\cdot 10^{-4}$ \\
0.18 & 800.00 & -497.76 & -8.45 & 44.3 & 98.0 & 1.3 & -6.11 & 18 & 23.5 &
6.0$\cdot 10^{-4}$ \\
0.20 & 445.80 & -443.30 & -8.82 & 51.8 & 97.8 & 1.7 & -5.96 & 19 & 25.0 &
6.0$\cdot 10^{-4}$ \\
0.25 & 0.00 & -388.67 & -8.99 & 74.6 & 97.0 & 2.0 & -5.36 & 20 & 25.0 &
5.9$\cdot 10^{-4}$ \\
0.30 & -71.50 & -394.50 & -8.02 & 115.0 & 95.0 & 3.0 & -4.20 & 21 & 27.0 & 
$4.0\cdot 10^{-4}$ \\
0.40 & -69.87 & -464.90 & -7.02 & 189.0 & 89.0 & 5.8 & -3.20 & 24 & 31.0 &
1.7$\cdot 10^{-4}$ \\
\hline
\end{tabular}

\end{table}

\begin{table}[h]
\caption[Parameters]{Parameters of the Moscow $NN$ potentials.}

\medskip

\begin{tabular}{|c|c|r|c|c|c|c|c|c|c|c|}
\hline
variant&\multicolumn{4}{|c|}{parameter
values}&\multicolumn{6}{|c|}{parameter values of $r_0$ [fm], $\lambda$ [MeV]}\\
\cline{2-5}
& $n$ & $V_0$ [MeV]& $\eta $[fm$^{-2}$] & $\alpha$[fm$^{-1}$] 
& \multicolumn{6}{|c|}{for particular channels}\\ \hline
 \multicolumn{5}{|c|}{triplet channels}& \multicolumn{2}{|c|}{$^3S_1$}&
 \multicolumn{2}{|c|}{$^3D_2$} &\multicolumn{2}{|c|}{$^3D_3$}\\ \hline
A & 3 & -466.7 & 1.600 & 3.000 & 0.448 & $\infty$ & 0.4600 & 251.0 & 0.300 &
$\infty$ \\
B & 3 & -1329.0 & 2.296 & 1.884 & 0.445 & $\infty$ & 0.4061 & 427.2 
& 0.276 & $\infty$ \\
%C & 5 & -1414.7 & 2.313 & 2.35 & 0.458 & $\infty$ & 0.4189 & 370.8 & 0.2919
%& $\infty$ \\
C & 7 & -1459.2 & 2.372 & 2.610 & 0.454 & $\infty$ & 0.4199 & 353.8 & 0.289
& $\infty$ \\ \hline \hline
 \multicolumn{5}{|c|}{singlet channels}& \multicolumn{2}{|c|}{$^1S_0$}&
 \multicolumn{2}{|c|}{$^1D_2$} & \multicolumn{2}{|c|}{ }\\ \hline
a & & -1106.2 & 1.600 & 3.000 & 0.4998 & $\infty$ & 0.4472 & 229.0 &
\multicolumn{2}{|c|}{ }\\
b & & -1220.0 & 1.753 & 1.884 & 0.4815 & $\infty$ & 0.4103 & 303.7 &
%\multicolumn{2}{|c|}{ }\\
%c & & -1214.0 & 1.754 & 3.00 & 0.4815 & $\infty$ & 0.4103 & 303.7 &
\multicolumn{2}{|c|}{ }\\
c & & -1222.0 & 1.738 & 1.000 & 0.4825 & $\infty$ & 0.4071 & 321.8 &
\multicolumn{2}{|c|}{ }\\ \hline
\end{tabular}
\bigskip
\end{table}

\begin{table}[h]

\caption[Deuteron] {Deuteron predictions for the different 
versions of the triplet-even Moscow NN potentials. $D_{loop}$ is 
the amplitude of the $D$-wave loop. The deuteron matter radius is defined as 
$r_m^2=1/4 \int_0^{\infty} \! dr  r^2 (u^2(r) + w^2(r) )$. 
For the most recent
value of the deuteron matter radius see Ref.\cite{Buc96}. Here, $\eta$ is the 
asymptotic $D/S$ ratio. In impulse approximation (model B) 
the deuteron magnetic moment is $\mu_d= 0.847 \, \mu_N$, 
the quadrupole moment is $Q_d= 0.271$ fm$^2$, and the 
charge radius is $r_{ch}= 2.112 \, $ fm. }
\bigskip
\begin{tabular}{|c|c|c|c|c|c|c|} \hline
Variant & $\chi ^2$ & $D_{loop}$ & $E_d$ [MeV] & $P_D\% $ &
$\eta $ & $r_m$ [fm] \\
\hline
A & 118 & 0.50 & -2.2244 & 6.59 & 0.0267 & 1.96 \\
B & 56 & 0.30 & -2.2246 & 5.75 & 0.0258 & 1.95 \\
%C & 94 & 0.07 & -2.2246 & 5.99 & 0.0261 & 1.94 \\
C & 126 & 0.02 & -2.2246 & 6.14 & 0.0262 & 1.94 \\
\hline
\end{tabular}
\bigskip
\end{table}

\begin{table}[h]

\caption[3N properties]{Three-nucleon properties with the A, B, 
C-versions of the Moscow potential.}
\bigskip
\begin{tabular}{|c|c|c|c|c|c|c|c|c|c|c|} \hline
variant of & $E(^3H)$ & $T(^3H)$ & $P_D$ & $P_P$ & $r_{ch}(^3H)$ &
$\Delta _{Coul}$ & $r_{ch}(^3He)$ & $q^2_{min}$ & $q^2_{max}$ &
$F_{ch}(^3H)_{max}$ \\
potential & \mbox{MeV} & \mbox{MeV} & $\% $ & $\% $ & \mbox{fm} &
\mbox{MeV} & \mbox{fm} & \mbox{fm}$^{-2}$ & \mbox{fm}$^{-2}$ & \\
\hline
A & -5.07 & 143.7 & 7.8 & 0.03 & 2.15 & 0.612 & 2.41 & 13 & 18 &
1.2$\cdot 10^{-3}$ \\
B & -6.03 & 162.4 & 7.4 & 0.03 & 2.01 & 0.670 & 2.24 & 16 & 22 &
9.6$\cdot 10^{-4}$ \\
C & -6.03 & 162.8 & 8.1 & 0.05 & 2.01 & 0.667 & 2.25 & 16 & 22 &
9.6$\cdot 10^{-4}$ \\ \hline
extrapol. & -8.48 & 175.1 & 8.5 & 0.06 & 1.88 & 0.737 & 2.07 & 17 & 23 &
$1.0\cdot 10^{-3}$ \\
to $E_{exp}$ &  &  &  &  &  &  &  &  &  &  \\ \hline
\mbox{experiment} & -8.48 &  &  &  & 1.76 & 0.764 & 1.96 & 13 & 18 &
$4\cdot 10^{-3}$ \\ \hline
\end{tabular}
\end{table}

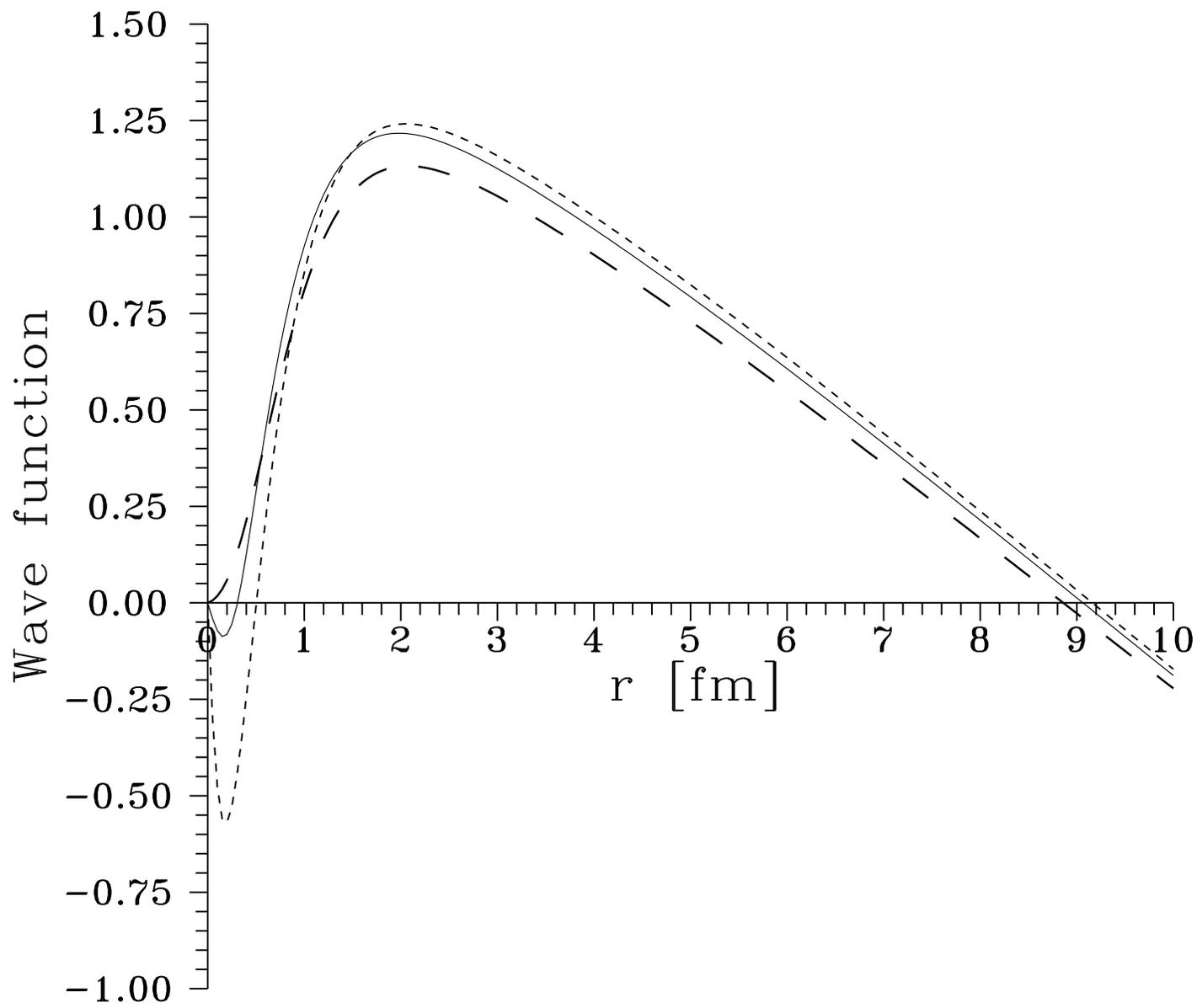
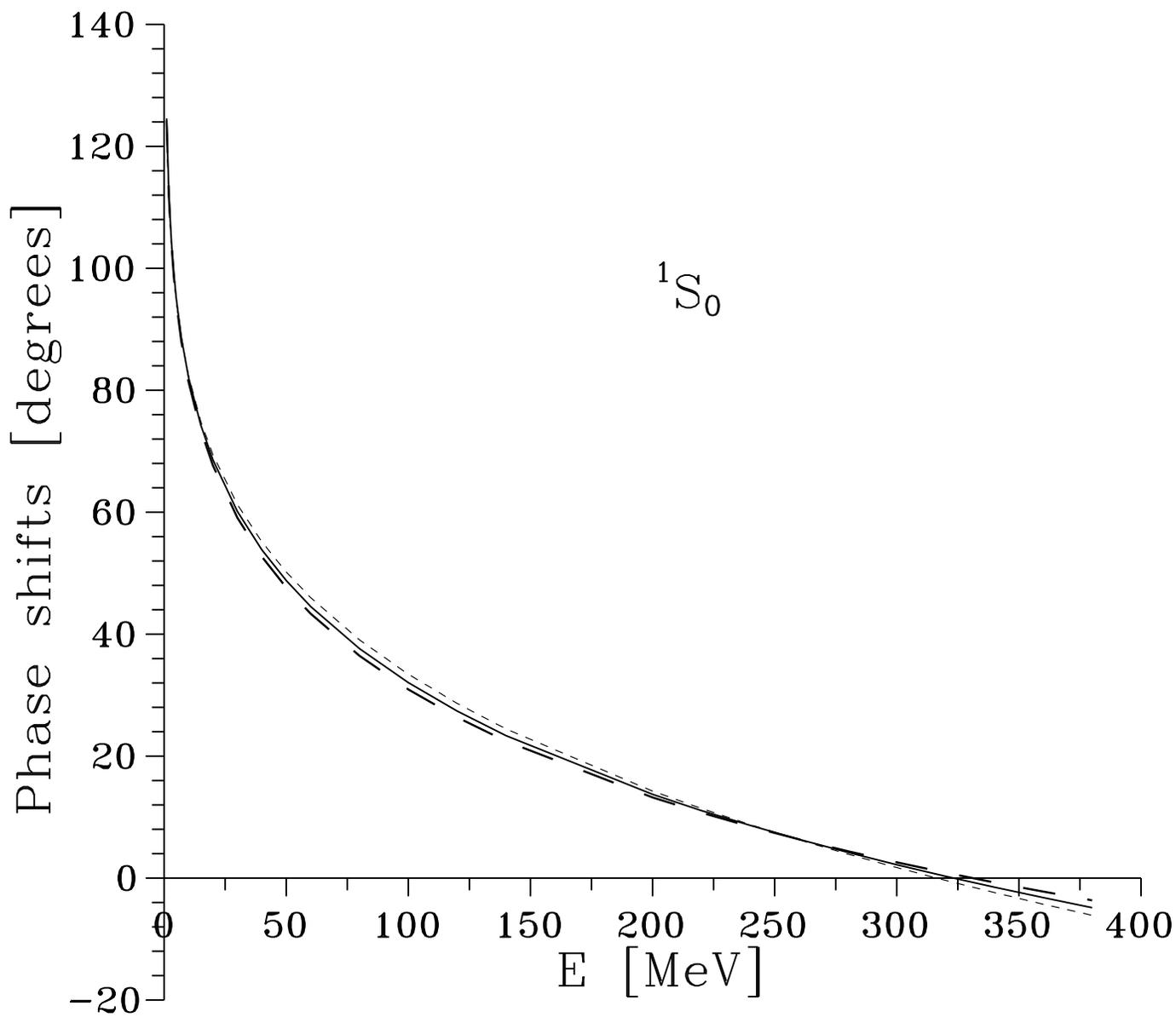
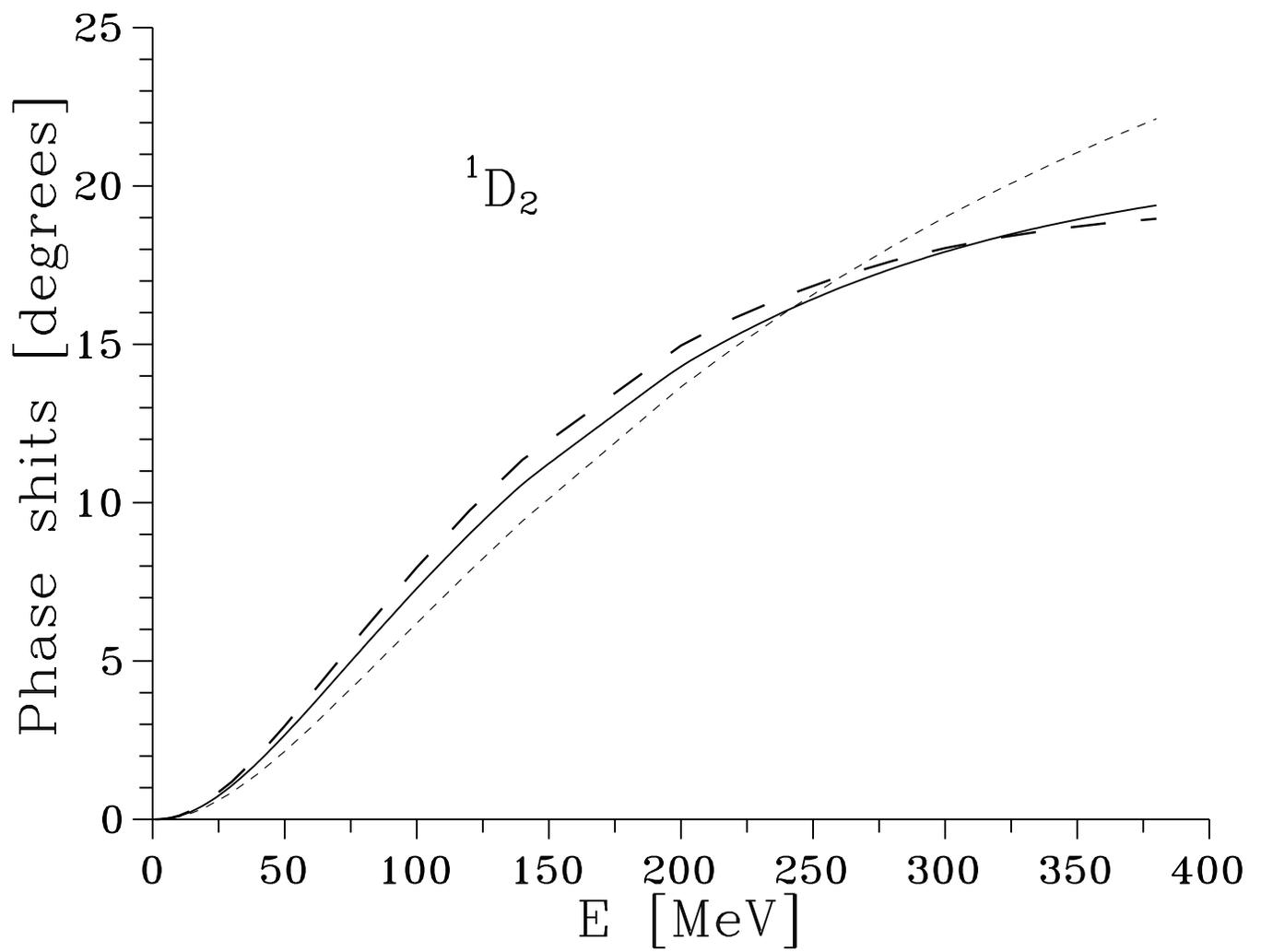
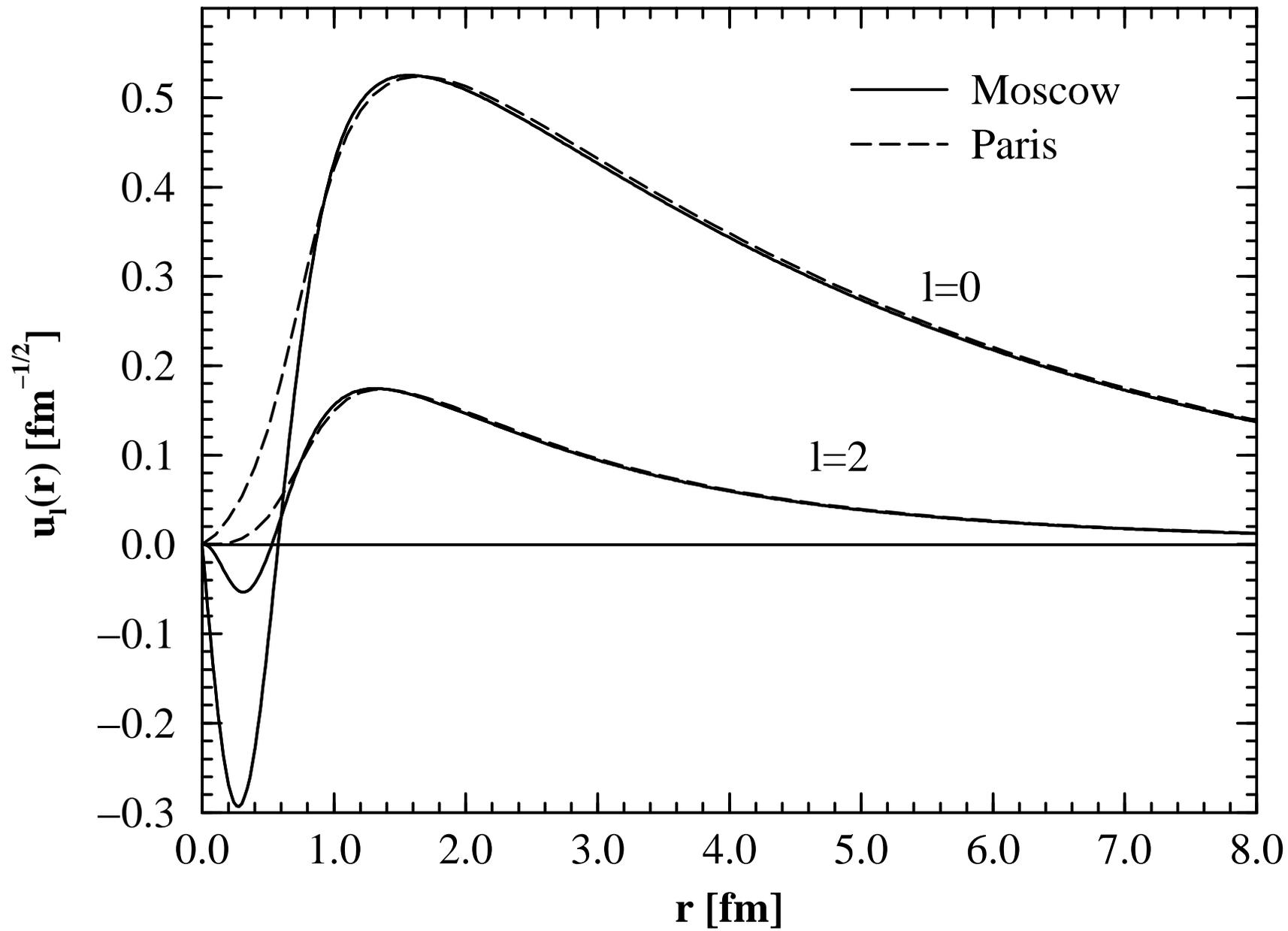
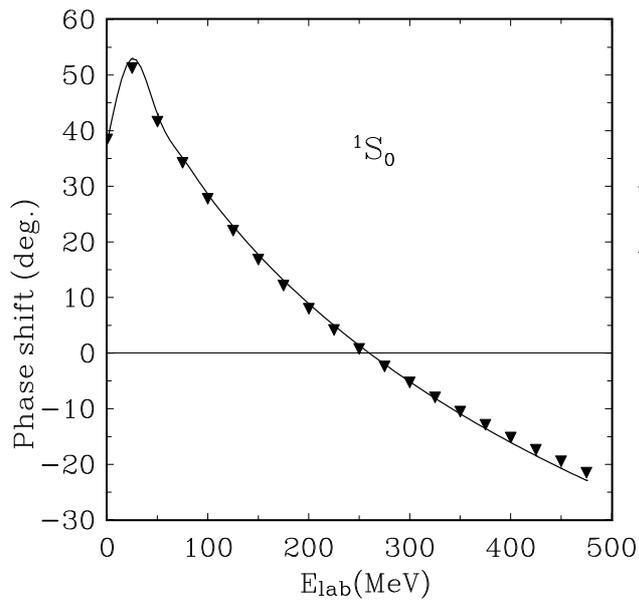
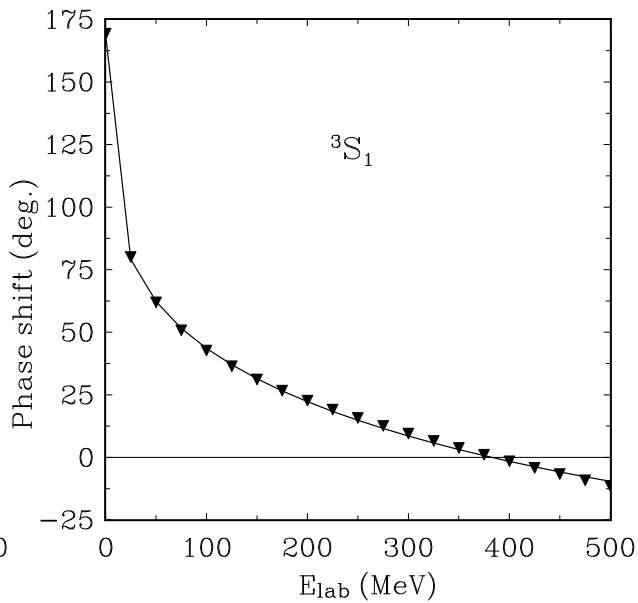
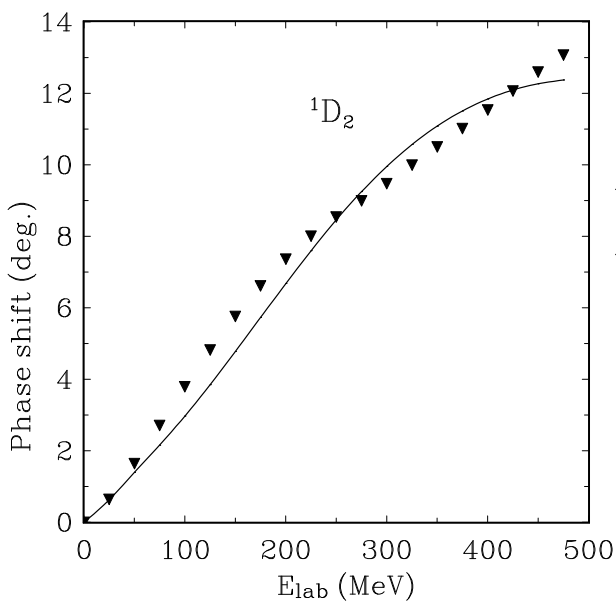
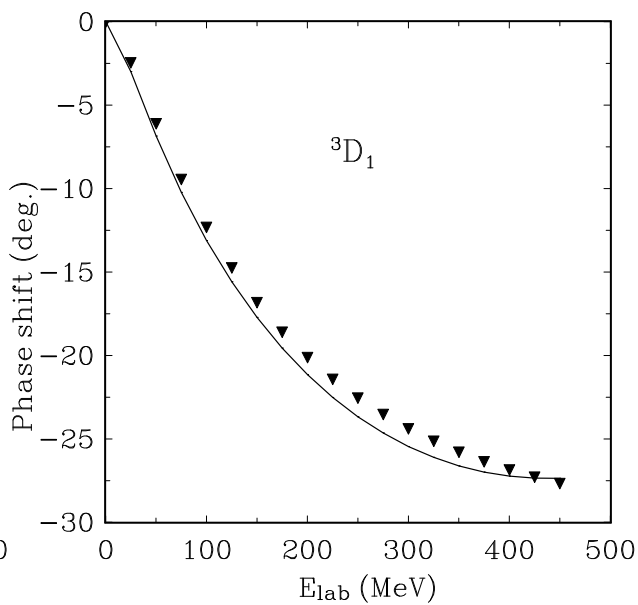
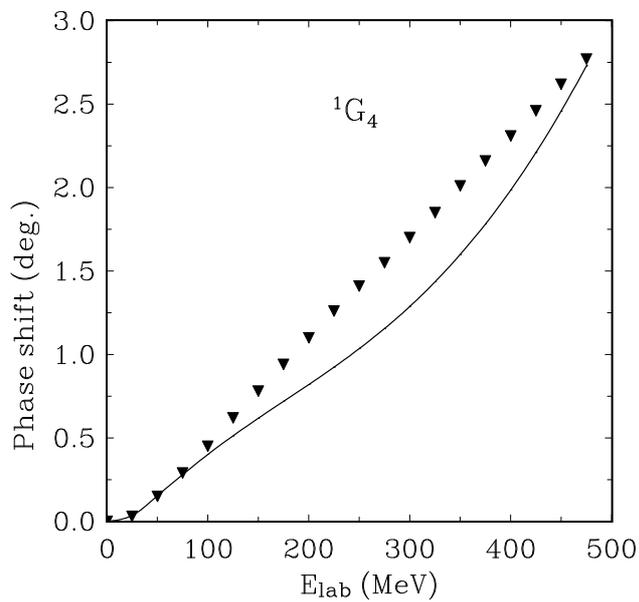
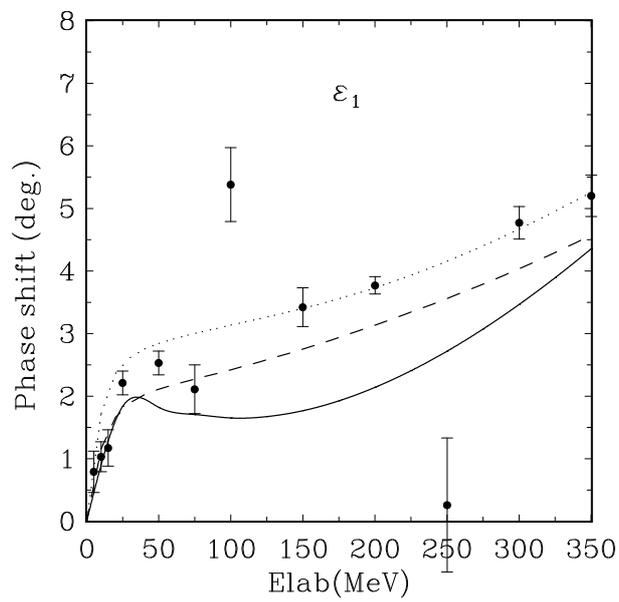
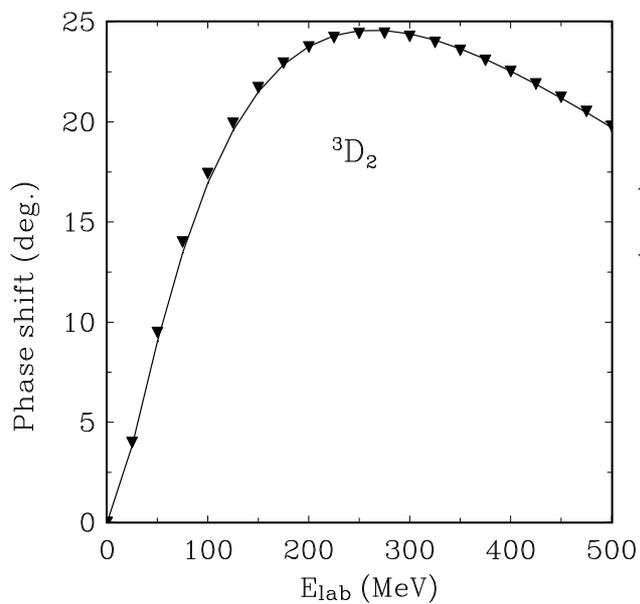
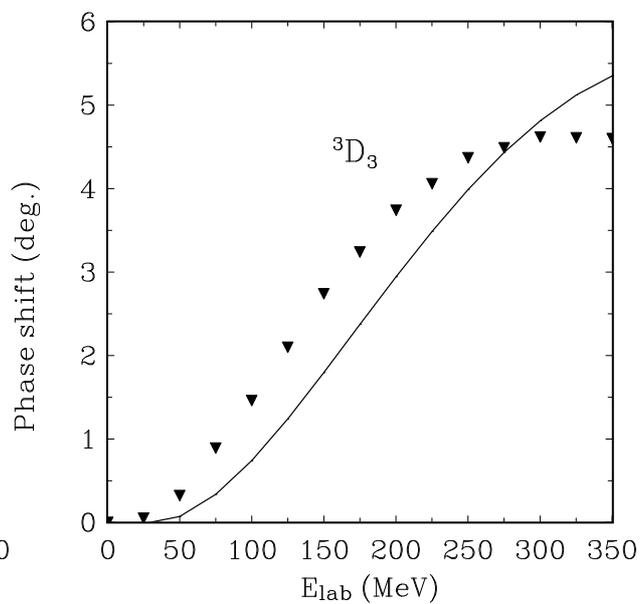
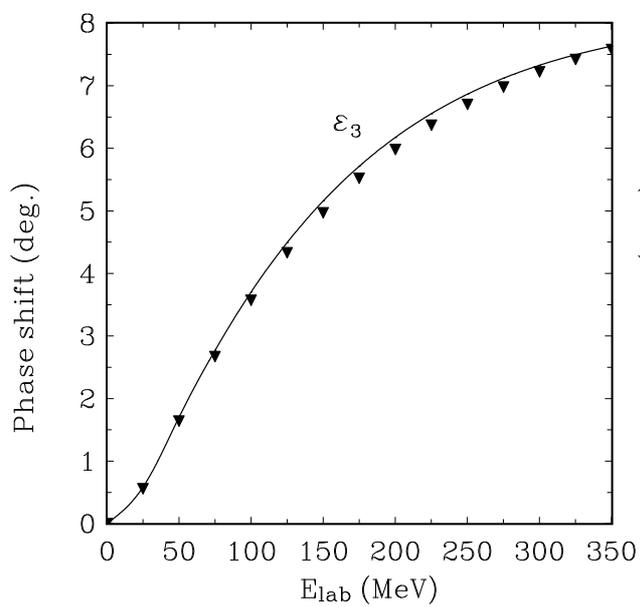
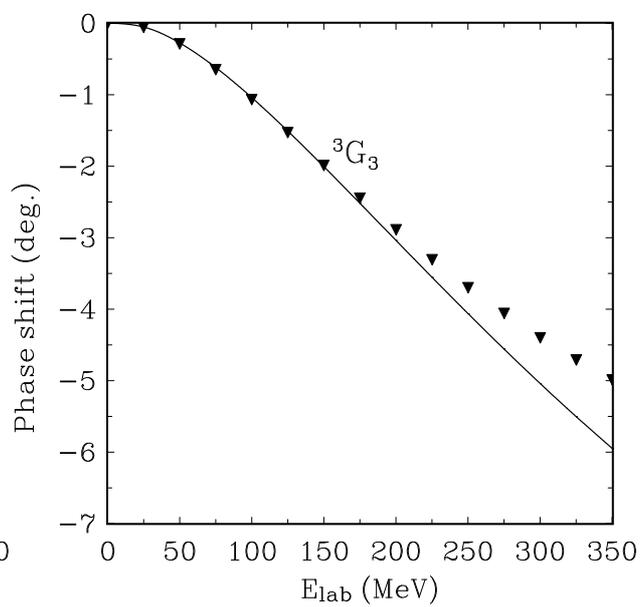
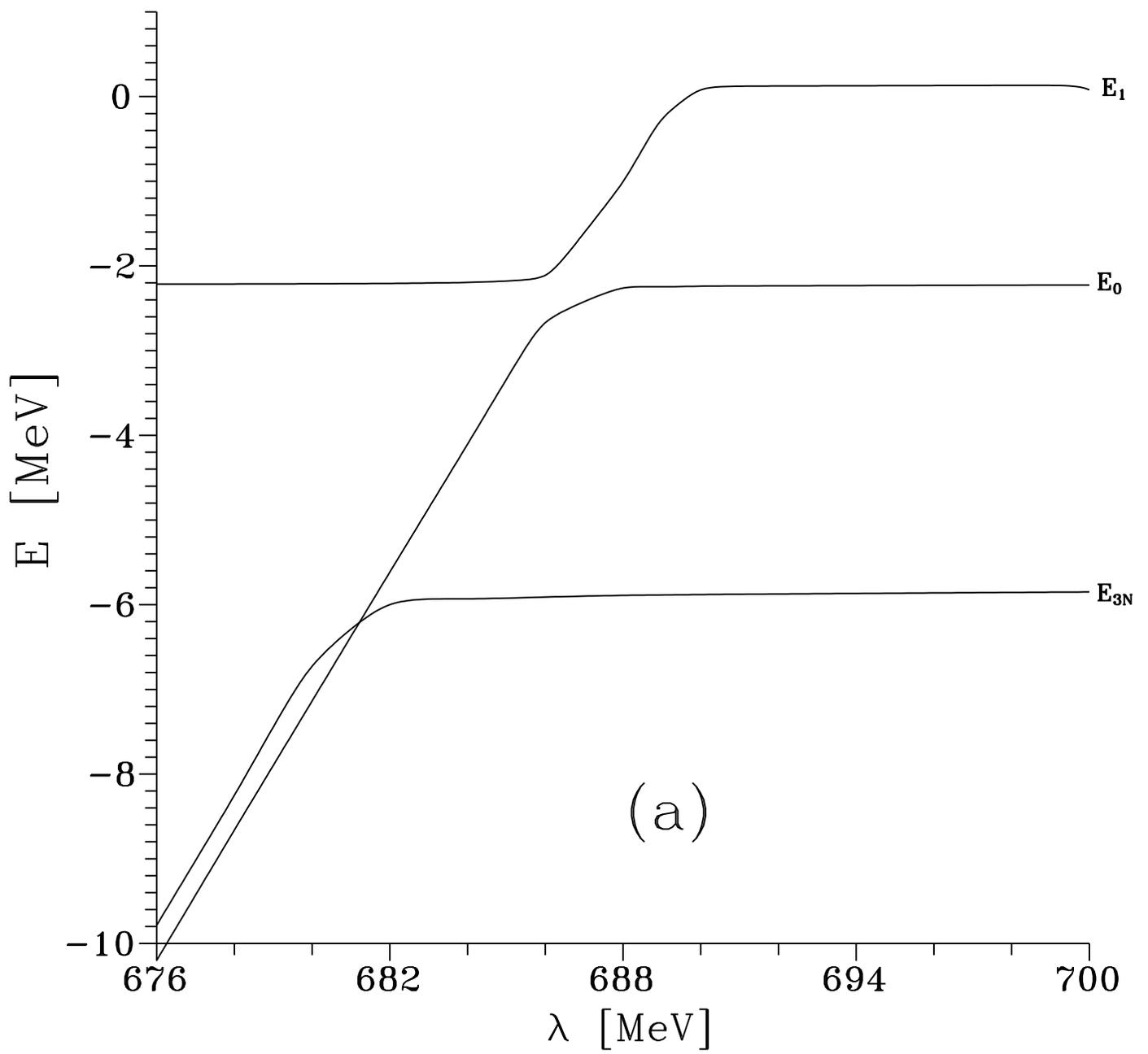
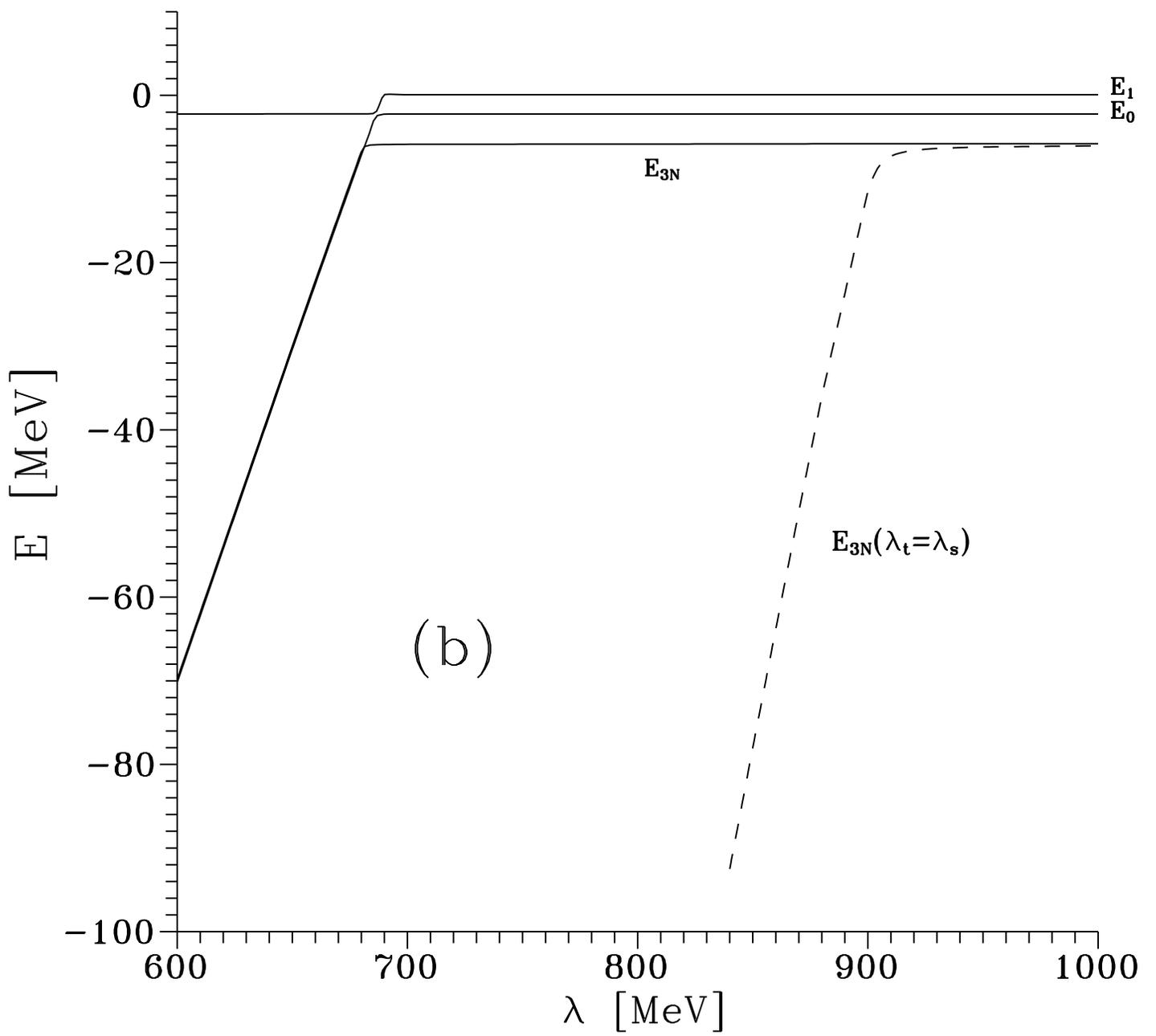
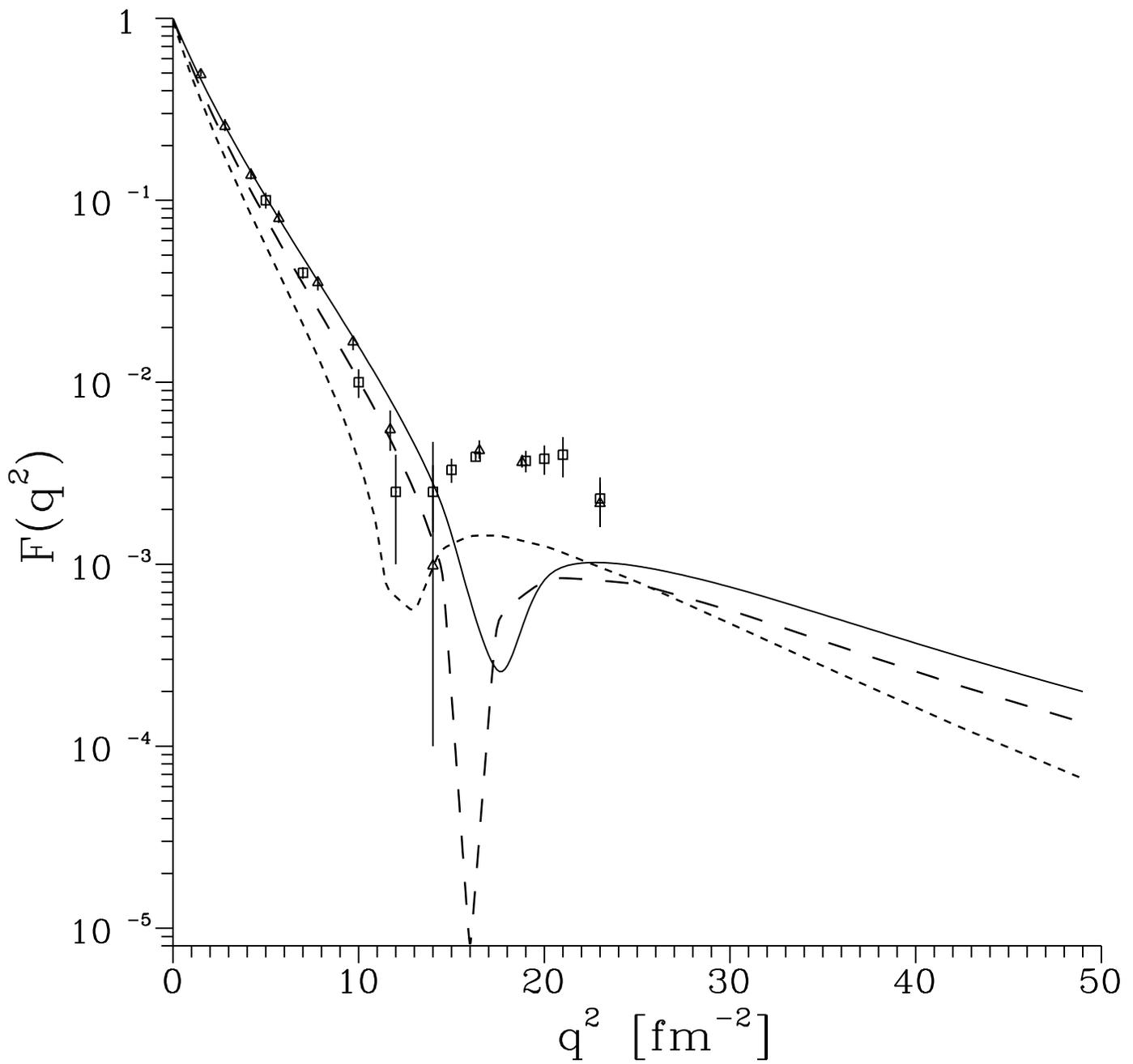
\end{document}